\pdfoutput=1
\documentclass[aps,10pt, prl,superscriptaddress,floatfix,twocolumn,showpacs,longbibliography]{revtex4-2}
\usepackage[colorlinks=true,linkcolor=blue,citecolor=blue]{hyperref}
\usepackage{amsfonts}
\usepackage{amsmath}
\usepackage{mathrsfs}
\usepackage{amssymb}
\usepackage{amsbsy}
\usepackage{epsfig}
\usepackage{graphicx}
\usepackage{epstopdf}
\usepackage{mathdots}
\usepackage{color}
\usepackage{cleveref}
\usepackage{physics}
\usepackage{bm}
\makeatletter

\newcommand{\Rmnum}[1]{\expandafter\@slowromancap\romannumeral #1@}
\makeatother

\begin{document} 
\title{Non-Hermitian higher-order topological insulators enabled by altermagnet engineering}

\author{Xiang Ji}
\affiliation{Department of Physics, Jiangsu University, Zhenjiang, 212013, China}
\affiliation{Eastern Institute for Advanced Study, Eastern Institute of Technology, Ningbo, Zhejiang 315200, China}
\affiliation{School of Physics and Astronomy, Shanghai Jiao Tong University, Shanghai 200240, China}

\author{Dengfeng Wang}
\affiliation{Department of Physics, Jiangsu University, Zhenjiang, 212013, China}

\author{Tong Zhou}
\altaffiliation{tzhou@eitech.edu.cn}
\affiliation{Eastern Institute for Advanced Study, Eastern Institute of Technology, Ningbo, Zhejiang 315200, China}

\author{Xiaosen Yang} \altaffiliation{yangxs@ujs.edu.cn}
\affiliation{Department of Physics, Jiangsu University, Zhenjiang, 212013, China}
\date{\today}

\begin{abstract}
We show that proximity to an altermagnet provides an efficient route to engineering non-Hermitian higher-order topological phases. The proximity-induced altermagnetic order gaps the edge states of a topological insulator, thereby driving a transition from a first-order to a second-order topological phase.
When combined with nonreciprocal hopping, the system exhibits both the non-Hermitian skin effect and a hybrid skin–topological effect, whereby first-order edge states and second-order corner states accumulate at selected corners of the lattice. We demonstrate that the spectral winding number of the edge states under cylindrical geometry dictates this corner localization and can be reversed by tuning the altermagnetic order. Consequently, both edge and corner states become directionally controllable. Our results establish altermagnets as a versatile platform for realizing and tuning skin-topological phenomena in non-Hermitian higher-order topological systems.
\end{abstract}

\maketitle

\section{Introduction}
Altermagnetism has recently emerged as a distinct class of collinear magnetic order that bridges the physics of ferromagnets and antiferromagnets~\cite{jpsj_2019hiroaki, prb2020alex, Libor_sciadv_2020, prx2022tomas, prx2022tomas2, afm2024yao_altermagnetism, songAltermagnetsNewClass2025a,jiangEnumerationSpinspaceGroups2024,PhysRevX2024xbchen,PhysRevX2024zyxiao, mcclartyLandauTheoryAltermagnetism2024, tongzhou_prl2025a, zhu2025twodimensional,song2025altermagnets, zhang2025crystalsymmetrypaired}.
It exhibits large momentum-dependent spin polarization while maintaining zero net magnetization, thus realizing time-reversal–odd spin splitting with compensated magnetic moments~\cite{tongzhou_prl2025a, zhang2025crystalsymmetrypaired}.
This unique symmetry realization enriches the taxonomy of magnetic phases and provides an intrinsic route to dissipationless spin control without stray fields. 
Its symmetry-filtered spin texture enables precise band-structure engineering, as verified theoretically and experimentally in
MnTe~\cite{aminNanoscaleImagingControl2024, gonzalezbetancourt2024anisotropic, PhysRevLett.132.036702}, CrSb~\cite{reimersDirectObservationAltermagnetic2024a, zhouManipulationAltermagneticOrder2025, yang2025threedimensional}, Mn$_5$Si$_3$~\cite{ PhysRevB.109.224430, badura2025observation}, RbV$_2$Se$_2$O~\cite{zhang2025crystalsymmetrypaired} and KV$_2$Se$_2$O~\cite{jiang2025metallic,wang2025atomicscalespinsensing2d}.
Due to its time-reversal breaking band structure and alternating momentum-dependent spin splitting, altermagnetism gives rise to a variety of novel effects and phases, including the crystal Hall effect~\cite{doi:10.1126/sciadv.aaz8809, sato2024altermagnetic, brekkeMinimalModelsTransport2024a, JinPhysRevLettSkyrmionHall}, giant tunneling magnetoresistance~\cite{Shao2021-ge, PhysRevX.12.011028}, spin–momentum interactions~\cite{reichlovaObservationSpontaneousAnomalous2024a, linCoulombDragAltermagnets2024a, ouassouDcJosephsonEffect2023a}, unique proximity effect~\cite{zhu2025altermagneticproximityeffect,zhu2025altermagnetoelectricspinfieldeffect} and topological phases~\cite{liMajoranaCornerModes2023a, ghorashiAltermagneticRoutesMajorana2024a, liCreationManipulationHigherorder2024a, liRealizingTunableHigherorder2024a, liFloatingEdgeBands2025}.

In recent years, the study of topological phases has been extended to non-Hermitian systems, attracting substantial research interest~\cite{rudner_topological_2009, hu_absence_2011, lee_anomalous_2016, shen_topological_2018, yao_edge_2018,yao_nonhermitian_2018, kunst_biorthogonal_2018, koutserimpas_nonreciprocal_2018, longhi_topological_2019, liu_secondorder_2019, luo_higherorder_2019,matsumoto_continuous_2020, xiao_nonhermitian_2020, bergholtz_exceptional_2021, ding_nonhermitian_2022, li_topological_2022, zhang_review_2022, weidemann_topological_2022, okuma_nonhermitian_2023,ji_generalized_2024,pnas2025_wbb,ji2025floqueta}.
Non-Hermitian systems exhibit a variety of striking phenomena that have no counterpart in Hermitian physics~\cite{bender_real_1998, hodaei_enhanced_2017, ashida_nonhermitian_2020, lin_topological_2023, wang_nonhermitian_2024}. Among the most prominent are exceptional points~\cite{miri_exceptional_2019,shuxuanprbep}, where both eigenvalues and eigenvectors coalesce, and the non-Hermitian skin effect (NHSE) \cite{yao_edge_2018,alvarez_nonhermitian_2018, song_nonhermitian_2019a, song_nonhermitian_2019, zhang_correspondence_2020, okuma_topological_2020, li_critical_2020, zhang_universal_2022, prl2024norio076502, zhaoTwodimensionalNonHermitianSkin2025a}, which drives an extensive number of bulk states to accumulate at system boundaries under open boundary conditions(OBC).
By reshaping the topology of complex energy spectra, non-Hermiticity enables unconventional topological phenomena, especially in higher dimensions. In particular, the hybridization of topological edge states with the NHSE gives rise to the skin–topological effect, where topologically protected edge states localize directionally at lower-dimensional boundaries~\cite{lee_hybrid_2019, li_topological_2020, kawabata_higherorder_2020, zou_observation_2021, fu_nonhermitian_2021, zhu_hybrid_2022, li_gainlossinduced_2022, sun_hybrid_2023, sun_photonic_2024,ji_nonhermitian_2024,JPCM2024_lhli}. 
These phenomena highlight the importance of understanding how non-Hermiticity interacts with intrinsic topological boundary states.
Exploring this interplay in altermagnetic systems is especially compelling: while recent works have examined exceptional-point physics in non-Hermitian altermagnets~\cite{prb2025dash,prb2024reja}, the NHSE and its coupling to higher-order topology in altermagnetic platforms remain largely unexplored.

In this work, we demonstrate that proximitizing an altermagnet to a non-Hermitian topological insulator provides an effective and controllable route to engineering non-Hermitian higher-order topological phases.
Altermagnetism opens a gap in the topological edge states of the topological insulator, leading to a topological phase transition from a first-order topological phase to a second-order topological phase.
When combined with nonreciprocal hopping, the system exhibits both the NHSE and the hybrid skin–topological effect.
The latter, manifested as the accumulation of first-order edge states at the corners of the systems, is dictated by the spectral winding number of the edge bands under cylindrical geometry.
Crucially, the proximitized altermagnetic order serves as a powerful tuning knob: by modifying the band dispersion of the edge states, it reverses the spectral winding number and thereby switches the localization direction of the topological edge states.
As the topological corner states inherit this skin–topological effect, they likewise accumulate at tunable corners of the lattice. These results establish altermagnetism as a versatile mechanism for designing and controlling non-Hermitian higher-order topological phases.

\section{Switchable Skin-topological effect}

\begin{figure*}
\centering     
\includegraphics[width=0.8\textwidth]{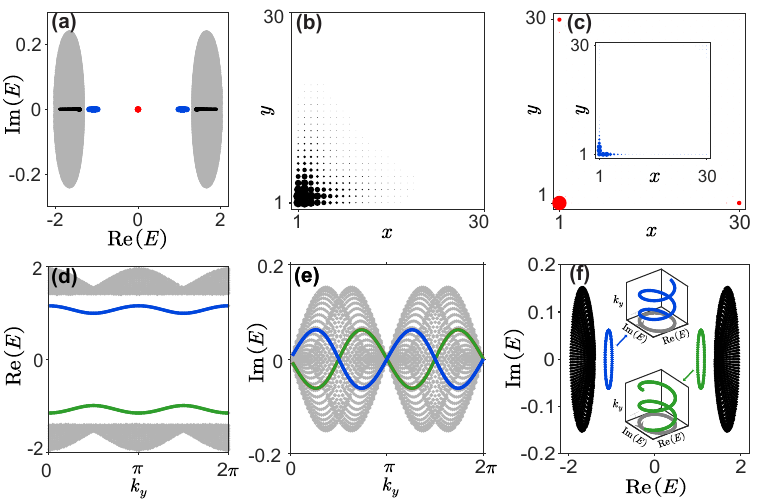}
\caption{Energy spectrum and spatial distributions in the presence of strong altermagnetism.
(a) Energy spectrum for a $30 \times 30$ square lattice under PBC (gray) and OBC (others), respectively.
(b) Average probability distributions of bulk states under OBC.
(c) Probability distributions of second-order corner states. 
    The inset shows the average probability distributions of first-order edge states.
(d) and (e) Real and imaginary parts of the energies as a function of $k_y$ for the system under OBC$_x$-PBC$_y$. Edge states with positive (negative) real energies are marked in blue (green).
(f) Complex energy spectrum of the system under OBC$_x$-PBC$_y$.
The insects show the spectrum winding of edge states.
Parameters: $t_x=t_y=A_x=A_y=1$, $\delta_x=\delta_y=0.2$, $J=0.7$, and $m_0=0$.}
\label{fig:fig1}
\end{figure*}

We consider a 2D non-Hermitian insulator in proximity to an altermagnet, whose Bloch Hamiltonian is given by
\begin{align}
H(\mathbf k) = H_\text{TI}(\mathbf k) + H_\text{NH}(\mathbf k) + H_\text{AM}(\mathbf k).
\end{align}
Here, $H_\text{TI}(\mathbf k)  = (m_0-t_x\cos k_x-t_y\cos k_y)\sigma_x+A_x \sin k_x \sigma_z s_x + A_y\sin k_y \sigma_z s_y$ describes a Hermitian topological insulator that preserves time-reversal symmetry and hosts helical edge states, relevant to materials like HgTe and WTe$_2$~\cite{science2006sczhang,yan_majorana_2019,science2014jli}.
The non-Hermiticity is introduced by $H_\text{NH}(\mathbf k)=(- i\delta_x\sin k_x - i\delta_y\sin k_y)\sigma_x$, which effectively models a nonreciprocal hopping arising from an imaginary gauge field.
The proximity-induced altermagnetic order~\cite{ghorashi_altermagnetic_2024,zhu2025altermagneticproximityeffect}, supported by multiple realistic materials, is given by $H_\text{AM}(\mathbf k)=J(\cos k_x-\cos k_y)\sigma_z s_z$, characterizing a $d$-wave spin splitting where $J$ denotes the strength of the altermagnetism. 
$s_i$ and $\sigma_i$ denote the Pauli matrices acting on spin and orbital degrees of freedom, respectively.

In the absence of altermagnetism ($J=0$), the Hamiltonian reduces to a non-Hermitian topological insulator hosting gapless helical edge states which is evenly localized at four edges of the lattice.
Turning on the altermagnetic order fundamentally restructures the phase diagram and drives the system into a higher-order topological phase.
In particular, we show the energy spectrum of the system with $J = 0.7$ in Fig.~\ref{fig:fig1}(a), where the spectrum under periodic boundary conditions (PBC) appears in gray, while the OBC spectrum are separated into three parts corresponding to bulk states (black), first-order edge states (blue), and second-order corner states (red).
The altermagnetic order gaps the helical edge states, generating four in-gap corner states and driving the system into a second-order topological insulator.
The nonreciprocal hopping renders the spectrum highly sensitive to the boundary conditions: relative to the PBC bands, the OBC bulk energies undergo a pronounced collapse, signaling the NHSE.
The spatial distribution in Fig.~\ref{fig:fig1}(b) confirms that the bulk states accumulate at the lower-left corner of the square lattice, consistent with the direction of nonreciprocal pumping.
Beyond NHSE of the bulk states, the interplay between altermagnet and non-Hermiticity also reshapes the localization of topological edge states.
As shown in the inset of Fig.~\ref{fig:fig1}(c), altermagnetic order and non-Hermiticity drives the first-order edge states to accumulate at the lower-left corner of the lattice, a hallmark of the skin–topological effect, in sharp contrast to the $J=0$ case  where the edge states remain uniformly distributed along the boundaries.
The skin–topological effect further influences the second-order corner states: as illustrated in Fig.~\ref{fig:fig1}(c), the in-gap zero states preferentially localize at the lower-left corner, rather than appearing symmetrically at all four corners as in the Hermitian limit or absence of altermagnetism. 

To gain a deeper understanding of the skin–topological states, we analyze the spctrum winding number of the edge states under a cylindrical geometry.
While the point-gap topology is the standard diagnostic of the NHSE in one-dimensional non-Hermitian systems~\cite{zhang_correspondence_2020,okuma_topological_2020}, here we use spectrum  to characterize the skin–topological effect.
Figure~\ref{fig:fig1}(d) and~\ref{fig:fig1}(e) show the real and imaginary parts of the spectrum for OBC along the $x$ direction and PBC along the $y$ direction (OBC$_x$-PBC$_y$), respectively.
As the edge states under this cylindrical geometry disperse both in their real and imaginary parts, their trajectories in the complex-energy plane form a point gap, as illustrated in the insets of Fig.~\ref{fig:fig1}(f).
The associated spectral winding number is defined as
\begin{align}
	\mathcal{W}(E)&=\oint_{0}^{2\pi}\frac{\dd k_y}{2\pi i}\frac{\partial}{\partial k_y}\log \det [H(k_y)-E], \nonumber\\
        &=\sum_{n=1}^{N}\oint_{0}^{2\pi}\frac{\dd k_y}{2\pi i}\frac{\partial}{\partial k_y} \arg [H(k_y)-E],
\end{align}
where $H(k_y)$ denotes the Hamiltonian of the system under OBC$_x$-PBC$_y$. $E$ is the reference energy chosen from the edge spectrum under full OBC, and counterclockwise rotation in the complex plane is taken as positive.
The complex energies of the edge states under OBC$_x$-PBC$_y$ are shown in Fig.~\ref{fig:fig1}(f).
We find that the spectral winding number is $+2$, indicating that the first-order edge states under full OBC accumulate at the lower-left corner of the lattice.
Notably, reversing the direction of nonreciprocal pumping, e.g., $(\delta\rightarrow -\delta)$, reverses the imaginary dispersion and changes the winding number to $-2$, causing the first-order edge states to accumulate at the upper-right corner.

\begin{figure}
\centering     
\includegraphics[width=0.45\textwidth]{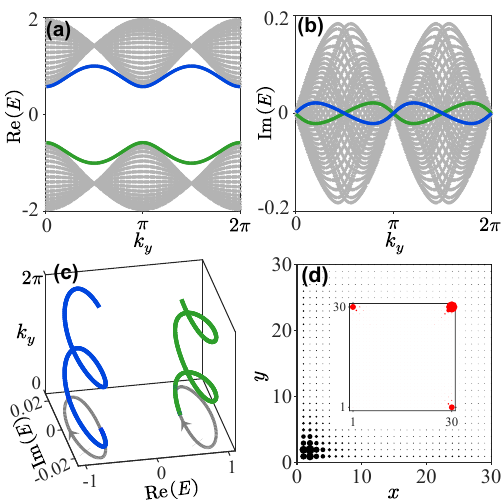}
\caption{Weak altermagnetism case ($J=0.3$).
(a) and (b) Real and imaginary parts of the energies as functions of $k_y$ for the system under cylindrical geometry. Topological edge states with positive (negative) real energies are marked in blue (green).
(c) Complex energy spectrum of the edge states as a function of $k_y$.
(d) Average probability distributions of bulk states under OBC. 
Inset: average probability distribution of zero states.}
\label{fig:fig2}
\end{figure}

Furthermore, varying the altermagnetic strength $J$ significantly modifies the real part of the edge-state dispersion under cylindrical geometry.
The localization behavior of the skin-topological states and second-order corner states changes when a relatively small $J$ is applied.
To illustrate this, we show the real and imaginary energies of the system under cylindrical geometry with $J = 0.3$ as a function of $k_y$ in Fig.~\ref{fig:fig2}(a) and~\ref{fig:fig2}(b), respectively.
The eigenvalues whose real parts are positive (negative) and closest to zero are marked in blue (green).
The sign of the real dispersion is opposite to the strong altermagnetism case in Fig.~\ref{fig:fig1}(d), while the signs of the imaginary dispersion remain the same.
Therefore, the spectral winding numbers under cylindrical geometry are reversed compared to the case of $J = 0.7$.
As shown in Fig.~\ref{fig:fig2}(c), the spectral winding numbers become $-2$, indicating that the first-order edge states will accumulate at the upper-right corner of the lattice, which is opposite to the strong altermagnetism case.
Affected by the skin-topological effect, although the bulk states localize at the lower-left corners of the system, the corner states tend to appear at the same corner as the first-order edge states and do not appear at the lower-left corner, as shown in Fig.~\ref{fig:fig2}(d).

Overall, the skin-topological states and second-order corner states are localized at the opposite corners of the lattice when we apply different strengths of altermagnetism, which can be well explained by the spectrum winding number of the edge states under cylindrical geometry.
Therefore, tuning the strength of the altermagnetism provides a feasible method to tune the localization behavior of the topological states.

\begin{figure}
\includegraphics[width=0.45\textwidth]{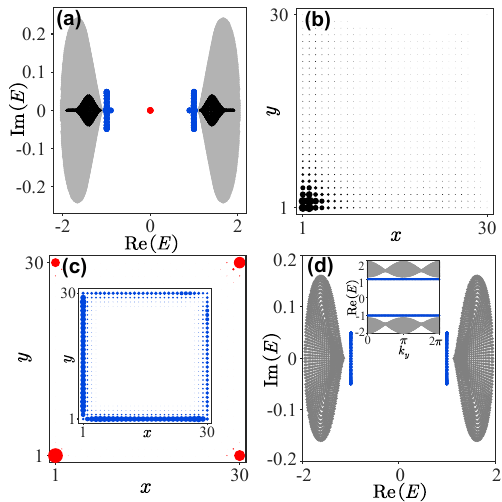}
\caption{The critical points with $J=0.57$. 
(a) Complex energy spectrum.
(b) Average probability distribution of the bulk states.
(c) Average probability distribution of the corner states and first-order edge states (inset) of the system under OBC.
(d) The complex spectrum of the system under OBC$_x$-PBC$_y$. The inset shows the real parts of energies as a function of $k_y$.
}\label{fig:fig3}
\end{figure}

\section{Edge flat bands and Bipolar skin-topological effect}
We have shown that both the skin-topological states and second-order corner states can be switched by tuning the altermagnetic order, implying the existence of a critical altermagnetic strength $J_{c}=0.57$ between the weak and strong regimes. 
To illustrate this transition, we present the energy spectrum and bulk states probability distribution in Fig.~\ref{fig:fig3}(a) and~\ref{fig:fig3}(b), respectively.
In this case, the system remains in a higher-order topological phase and hosts four in-gap zero states.
While the bulk states under OBC accumulate at the lower-left corners of the lattice due to of the nonreciprocal hopping. 
The first-order edge states are uniformly distributed along all four edges, as shown in the inset of Fig.~\ref{fig:fig3}(c), indicating the absence of the skin–topological effect, as shown in Fig.~\ref{fig:fig3}(c). The second-order corner states also appear symmetrically at all four corners without any directional preference. This absence of both edge localization and corner selectivity originates from the flat real energy dispersion of hte edge states under cylindrical geometry, as depicted in Fig.~\ref{fig:fig3}(d), which results in a zero spectral winding number and suppresses the skin–topological effect. 

\begin{figure}
\includegraphics[width=0.45\textwidth]{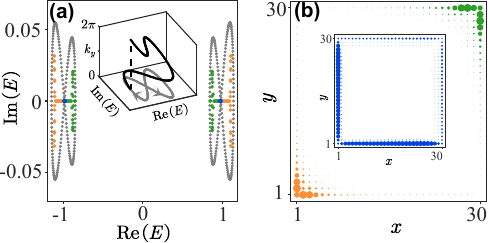}
\caption{The bipolar skin-topological effect. 
(a) The gray points depict the energies of the edge states under OBC$_x$-PBC$_y$, while colorful points represent the energies under full OBC with $m_0=0.2$.
The inset shows the complex spectrum under cylindrical geometry as a function of $k_y$.
(b) Average probability distribution of the edge states under full OBC, with colors corresponding to the energies highlighted in (a).
}\label{fig:fig4}
\end{figure}

Although our analysis primarily focuses on the $m_0 = 0$ case, the underlying control mechanism remains robust for finite $m_0$, demonstrating the physical universality of the observed phenomena across a broader parameter regime.
When a finite $m_0$ is introduced, the winding behavior of the edge states under cylindrical geometry becomes more intricate, and a bipolar skin–topological effect can emerge near the transition point discussed above.
As an illustrative example, we take $m_0 = 0.2$ and $J = 0.57$, and plot the spectrum of the first-order edge states in Fig.~\ref{fig:fig4}(a).
The energies obtained under full OBC and cylindrical geometry are shown as gray and colored points, respectively.
The edge states under full OBC with spectral winding numbers $+1$ and $-1$, as extracted from the inset, are marked in orange and green, respectively.
As shown in Fig.~\ref{fig:fig4}(b), these states localize at opposite corners of the lattice, with the probability distributions color-matched to the corresponding points in Fig.~\ref{fig:fig4}(a).
Remarkably, the states at the intersections between regions of opposite winding numbers correspond to Bloch-edge states, marked by blue points in Fig.~\ref{fig:fig4}(a). These states are extended along the edges rather than localized at the corners, as confirmed in the inset of Fig.~\ref{fig:fig4}(b).

\section{Conclusion}
In this work, we have investigated the non-Hermitian second-order topological insulator realized by placing a non-Hermitian topological insulator in proximity to an altermagnet.
The nonreciprocal hopping drives the bulk states to accumulate at the lattice corners, generating a pronounced non-Hermitian skin effect.
Meanwhile, the altermagnetic order gaps the topological edge states and induces a transition from a first-order to a second-order topological phase.
The interplay between altermagnetism and non-Hermiticity produces a skin–topological effect, in which the first-order edge states are driven to the corners of the lattice.
This effect is accurately captured by the spectral winding number of the edge states under cylindrical geometry.
Crucially, the proximitized altermagnetic order reshapes the real-part edge dispersion, enabling a sign reversal of the spectral winding number and thus a controllable inversion of edge and corner localization, without altering the direction of nonreciprocal hopping.
As a consequence, the second-order corner states inherit the skin–topological localization and preferentially accumulate at the same corners as the edge states.
Our results reveal the rich interplay between altermagnetism and non-Hermiticity, demonstrating that non-Hermitian altermagnets offer a versatile and tunable platform for manipulating topological edge and corner states.

The nonreciprocal hopping in our model can be readily implemented in cold-atom platforms using Raman-assisted schemes and programmable lattices~\cite{liang_dynamic_2022}, acoustics~\cite{weiweiPhysRevLettacous}, topological circuits~\cite{helbig_generalized_2020}, and optics~\cite{xiao_nonhermitian_2020}.
In electronic settings, nonreciprocal hopping may be effectively engineered via bilayer or multi-orbital coupling.
Although we characterize the skin–topological effect using the spectral winding number under cylindrical geometry, a comprehensive understanding of such effects across different dimensions and symmetry classes remains to be developed.
Furthermore, the interaction between skin–topological states and higher-order topological states represents an open question that merits further exploration.

\section{Acknowledgements}
This work was supported by Natural Science Foundation of Jiangsu Province (Grant No. BK20231320), National Natural Science Foundation of China (12474155) and the Zhejiang Provincial Natural Science Foundation of China (LR25A040001).

\appendix

\bibliography{reference} 

\begin{thebibliography}{99}%
\makeatletter
\providecommand \@ifxundefined [1]{%
 \@ifx{#1\undefined}
}%
\providecommand \@ifnum [1]{%
 \ifnum #1\expandafter \@firstoftwo
 \else \expandafter \@secondoftwo
 \fi
}%
\providecommand \@ifx [1]{%
 \ifx #1\expandafter \@firstoftwo
 \else \expandafter \@secondoftwo
 \fi
}%
\providecommand \natexlab [1]{#1}%
\providecommand \enquote  [1]{``#1''}%
\providecommand \bibnamefont  [1]{#1}%
\providecommand \bibfnamefont [1]{#1}%
\providecommand \citenamefont [1]{#1}%
\providecommand \href@noop [0]{\@secondoftwo}%
\providecommand \href [0]{\begingroup \@sanitize@url \@href}%
\providecommand \@href[1]{\@@startlink{#1}\@@href}%
\providecommand \@@href[1]{\endgroup#1\@@endlink}%
\providecommand \@sanitize@url [0]{\catcode `\\12\catcode `\$12\catcode `\&12\catcode `\#12\catcode `\^12\catcode `\_12\catcode `\%12\relax}%
\providecommand \@@startlink[1]{}%
\providecommand \@@endlink[0]{}%
\providecommand \url  [0]{\begingroup\@sanitize@url \@url }%
\providecommand \@url [1]{\endgroup\@href {#1}{\urlprefix }}%
\providecommand \urlprefix  [0]{URL }%
\providecommand \Eprint [0]{\href }%
\providecommand \doibase [0]{https://doi.org/}%
\providecommand \selectlanguage [0]{\@gobble}%
\providecommand \bibinfo  [0]{\@secondoftwo}%
\providecommand \bibfield  [0]{\@secondoftwo}%
\providecommand \translation [1]{[#1]}%
\providecommand \BibitemOpen [0]{}%
\providecommand \bibitemStop [0]{}%
\providecommand \bibitemNoStop [0]{.\EOS\space}%
\providecommand \EOS [0]{\spacefactor3000\relax}%
\providecommand \BibitemShut  [1]{\csname bibitem#1\endcsname}%
\let\auto@bib@innerbib\@empty
\bibitem [{\citenamefont {Hayami}\ \emph {et~al.}(2019)\citenamefont {Hayami}, \citenamefont {Yanagi},\ and\ \citenamefont {Kusunose}}]{jpsj_2019hiroaki}%
  \BibitemOpen
  \bibfield  {author} {\bibinfo {author} {\bibfnamefont {S.}~\bibnamefont {Hayami}}, \bibinfo {author} {\bibfnamefont {Y.}~\bibnamefont {Yanagi}},\ and\ \bibinfo {author} {\bibfnamefont {H.}~\bibnamefont {Kusunose}},\ }\bibfield  {title} {\bibinfo {title} {Momentum-dependent spin splitting by collinear antiferromagnetic ordering},\ }\href {https://doi.org/10.7566/JPSJ.88.123702} {\bibfield  {journal} {\bibinfo  {journal} {J. Phys. Soc. Jpn.}\ }\textbf {\bibinfo {volume} {88}},\ \bibinfo {pages} {123702} (\bibinfo {year} {2019})}\BibitemShut {NoStop}%
\bibitem [{\citenamefont {Yuan}\ \emph {et~al.}(2020)\citenamefont {Yuan}, \citenamefont {Wang}, \citenamefont {Luo}, \citenamefont {Rashba},\ and\ \citenamefont {Zunger}}]{prb2020alex}%
  \BibitemOpen
  \bibfield  {author} {\bibinfo {author} {\bibfnamefont {L.-D.}\ \bibnamefont {Yuan}}, \bibinfo {author} {\bibfnamefont {Z.}~\bibnamefont {Wang}}, \bibinfo {author} {\bibfnamefont {J.-W.}\ \bibnamefont {Luo}}, \bibinfo {author} {\bibfnamefont {E.~I.}\ \bibnamefont {Rashba}},\ and\ \bibinfo {author} {\bibfnamefont {A.}~\bibnamefont {Zunger}},\ }\bibfield  {title} {\bibinfo {title} {Giant momentum-dependent spin splitting in centrosymmetric low-$z$ antiferromagnets},\ }\href {https://doi.org/10.1103/PhysRevB.102.014422} {\bibfield  {journal} {\bibinfo  {journal} {Phys. Rev. B}\ }\textbf {\bibinfo {volume} {102}},\ \bibinfo {pages} {014422} (\bibinfo {year} {2020})}\BibitemShut {NoStop}%
\bibitem [{\citenamefont {Šmejkal}\ \emph {et~al.}(2020)\citenamefont {Šmejkal}, \citenamefont {González-Hernández}, \citenamefont {Jungwirth},\ and\ \citenamefont {Sinova}}]{Libor_sciadv_2020}%
  \BibitemOpen
  \bibfield  {author} {\bibinfo {author} {\bibfnamefont {L.}~\bibnamefont {Šmejkal}}, \bibinfo {author} {\bibfnamefont {R.}~\bibnamefont {González-Hernández}}, \bibinfo {author} {\bibfnamefont {T.}~\bibnamefont {Jungwirth}},\ and\ \bibinfo {author} {\bibfnamefont {J.}~\bibnamefont {Sinova}},\ }\bibfield  {title} {\bibinfo {title} {Crystal time-reversal symmetry breaking and spontaneous hall effect in collinear antiferromagnets},\ }\href {https://doi.org/10.1126/sciadv.aaz8809} {\bibfield  {journal} {\bibinfo  {journal} {Sci. Adv.}\ }\textbf {\bibinfo {volume} {6}},\ \bibinfo {pages} {eaaz8809} (\bibinfo {year} {2020})}\BibitemShut {NoStop}%
\bibitem [{\citenamefont {\ifmmode~\check{S}\else \v{S}\fi{}mejkal}\ \emph {et~al.}(2022{\natexlab{a}})\citenamefont {\ifmmode~\check{S}\else \v{S}\fi{}mejkal}, \citenamefont {Sinova},\ and\ \citenamefont {Jungwirth}}]{prx2022tomas}%
  \BibitemOpen
  \bibfield  {author} {\bibinfo {author} {\bibfnamefont {L.}~\bibnamefont {\ifmmode~\check{S}\else \v{S}\fi{}mejkal}}, \bibinfo {author} {\bibfnamefont {J.}~\bibnamefont {Sinova}},\ and\ \bibinfo {author} {\bibfnamefont {T.}~\bibnamefont {Jungwirth}},\ }\bibfield  {title} {\bibinfo {title} {Beyond conventional ferromagnetism and antiferromagnetism: A phase with nonrelativistic spin and crystal rotation symmetry},\ }\href {https://doi.org/10.1103/PhysRevX.12.031042} {\bibfield  {journal} {\bibinfo  {journal} {Phys. Rev. X}\ }\textbf {\bibinfo {volume} {12}},\ \bibinfo {pages} {031042} (\bibinfo {year} {2022}{\natexlab{a}})}\BibitemShut {NoStop}%
\bibitem [{\citenamefont {\ifmmode~\check{S}\else \v{S}\fi{}mejkal}\ \emph {et~al.}(2022{\natexlab{b}})\citenamefont {\ifmmode~\check{S}\else \v{S}\fi{}mejkal}, \citenamefont {Sinova},\ and\ \citenamefont {Jungwirth}}]{prx2022tomas2}%
  \BibitemOpen
  \bibfield  {author} {\bibinfo {author} {\bibfnamefont {L.}~\bibnamefont {\ifmmode~\check{S}\else \v{S}\fi{}mejkal}}, \bibinfo {author} {\bibfnamefont {J.}~\bibnamefont {Sinova}},\ and\ \bibinfo {author} {\bibfnamefont {T.}~\bibnamefont {Jungwirth}},\ }\bibfield  {title} {\bibinfo {title} {Emerging research landscape of altermagnetism},\ }\href {https://doi.org/10.1103/PhysRevX.12.040501} {\bibfield  {journal} {\bibinfo  {journal} {Phys. Rev. X}\ }\textbf {\bibinfo {volume} {12}},\ \bibinfo {pages} {040501} (\bibinfo {year} {2022}{\natexlab{b}})}\BibitemShut {NoStop}%
\bibitem [{\citenamefont {Bai}\ \emph {et~al.}(2024)\citenamefont {Bai}, \citenamefont {Feng}, \citenamefont {Liu}, \citenamefont {Šmejkal}, \citenamefont {Mokrousov},\ and\ \citenamefont {Yao}}]{afm2024yao_altermagnetism}%
  \BibitemOpen
  \bibfield  {author} {\bibinfo {author} {\bibfnamefont {L.}~\bibnamefont {Bai}}, \bibinfo {author} {\bibfnamefont {W.}~\bibnamefont {Feng}}, \bibinfo {author} {\bibfnamefont {S.}~\bibnamefont {Liu}}, \bibinfo {author} {\bibfnamefont {L.}~\bibnamefont {Šmejkal}}, \bibinfo {author} {\bibfnamefont {Y.}~\bibnamefont {Mokrousov}},\ and\ \bibinfo {author} {\bibfnamefont {Y.}~\bibnamefont {Yao}},\ }\bibfield  {title} {\bibinfo {title} {Altermagnetism: Exploring new frontiers in magnetism and spintronics},\ }\href {https://doi.org/https://doi.org/10.1002/adfm.202409327} {\bibfield  {journal} {\bibinfo  {journal} {Adv. Funct. Mater.}\ }\textbf {\bibinfo {volume} {34}},\ \bibinfo {pages} {2409327} (\bibinfo {year} {2024})}\BibitemShut {NoStop}%
\bibitem [{\citenamefont {Song}\ \emph {et~al.}(2025{\natexlab{a}})\citenamefont {Song}, \citenamefont {Bai}, \citenamefont {Zhou}, \citenamefont {Han}, \citenamefont {Reichlova}, \citenamefont {Dil}, \citenamefont {Liu}, \citenamefont {Chen},\ and\ \citenamefont {Pan}}]{songAltermagnetsNewClass2025a}%
  \BibitemOpen
  \bibfield  {author} {\bibinfo {author} {\bibfnamefont {C.}~\bibnamefont {Song}}, \bibinfo {author} {\bibfnamefont {H.}~\bibnamefont {Bai}}, \bibinfo {author} {\bibfnamefont {Z.}~\bibnamefont {Zhou}}, \bibinfo {author} {\bibfnamefont {L.}~\bibnamefont {Han}}, \bibinfo {author} {\bibfnamefont {H.}~\bibnamefont {Reichlova}}, \bibinfo {author} {\bibfnamefont {J.~H.}\ \bibnamefont {Dil}}, \bibinfo {author} {\bibfnamefont {J.}~\bibnamefont {Liu}}, \bibinfo {author} {\bibfnamefont {X.}~\bibnamefont {Chen}},\ and\ \bibinfo {author} {\bibfnamefont {F.}~\bibnamefont {Pan}},\ }\bibfield  {title} {\bibinfo {title} {Altermagnets as a new class of functional materials},\ }\href {https://doi.org/10.1038/s41578-025-00779-1} {\bibfield  {journal} {\bibinfo  {journal} {Nat. Rev. Mater.}\ }\textbf {\bibinfo {volume} {10}},\ \bibinfo {pages} {473} (\bibinfo {year} {2025}{\natexlab{a}})}\BibitemShut {NoStop}%
\bibitem [{\citenamefont {Jiang}\ \emph {et~al.}(2024)\citenamefont {Jiang}, \citenamefont {Song}, \citenamefont {Zhu}, \citenamefont {Fang}, \citenamefont {Weng}, \citenamefont {Liu}, \citenamefont {Yang},\ and\ \citenamefont {Fang}}]{jiangEnumerationSpinspaceGroups2024}%
  \BibitemOpen
  \bibfield  {author} {\bibinfo {author} {\bibfnamefont {Y.}~\bibnamefont {Jiang}}, \bibinfo {author} {\bibfnamefont {Z.}~\bibnamefont {Song}}, \bibinfo {author} {\bibfnamefont {T.}~\bibnamefont {Zhu}}, \bibinfo {author} {\bibfnamefont {Z.}~\bibnamefont {Fang}}, \bibinfo {author} {\bibfnamefont {H.}~\bibnamefont {Weng}}, \bibinfo {author} {\bibfnamefont {Z.-X.}\ \bibnamefont {Liu}}, \bibinfo {author} {\bibfnamefont {J.}~\bibnamefont {Yang}},\ and\ \bibinfo {author} {\bibfnamefont {C.}~\bibnamefont {Fang}},\ }\bibfield  {title} {\bibinfo {title} {Enumeration of spin-space groups: Towards a complete description of symmetries of magnetic orders},\ }\href {https://doi.org/10.1103/PhysRevX.14.031039} {\bibfield  {journal} {\bibinfo  {journal} {Phys. Rev. X}\ }\textbf {\bibinfo {volume} {14}},\ \bibinfo {pages} {031039} (\bibinfo {year} {2024})}\BibitemShut {NoStop}%
\bibitem [{\citenamefont {Chen}\ \emph {et~al.}(2024)\citenamefont {Chen}, \citenamefont {Ren}, \citenamefont {Zhu}, \citenamefont {Yu}, \citenamefont {Zhang}, \citenamefont {Liu}, \citenamefont {Li}, \citenamefont {Liu}, \citenamefont {Li},\ and\ \citenamefont {Liu}}]{PhysRevX2024xbchen}%
  \BibitemOpen
  \bibfield  {author} {\bibinfo {author} {\bibfnamefont {X.}~\bibnamefont {Chen}}, \bibinfo {author} {\bibfnamefont {J.}~\bibnamefont {Ren}}, \bibinfo {author} {\bibfnamefont {Y.}~\bibnamefont {Zhu}}, \bibinfo {author} {\bibfnamefont {Y.}~\bibnamefont {Yu}}, \bibinfo {author} {\bibfnamefont {A.}~\bibnamefont {Zhang}}, \bibinfo {author} {\bibfnamefont {P.}~\bibnamefont {Liu}}, \bibinfo {author} {\bibfnamefont {J.}~\bibnamefont {Li}}, \bibinfo {author} {\bibfnamefont {Y.}~\bibnamefont {Liu}}, \bibinfo {author} {\bibfnamefont {C.}~\bibnamefont {Li}},\ and\ \bibinfo {author} {\bibfnamefont {Q.}~\bibnamefont {Liu}},\ }\bibfield  {title} {\bibinfo {title} {Enumeration and representation theory of spin space groups},\ }\href {https://doi.org/10.1103/PhysRevX.14.031038} {\bibfield  {journal} {\bibinfo  {journal} {Phys. Rev. X}\ }\textbf {\bibinfo {volume} {14}},\ \bibinfo {pages} {031038} (\bibinfo {year} {2024})}\BibitemShut {NoStop}%
\bibitem [{\citenamefont {Xiao}\ \emph {et~al.}(2024)\citenamefont {Xiao}, \citenamefont {Zhao}, \citenamefont {Li}, \citenamefont {Shindou},\ and\ \citenamefont {Song}}]{PhysRevX2024zyxiao}%
  \BibitemOpen
  \bibfield  {author} {\bibinfo {author} {\bibfnamefont {Z.}~\bibnamefont {Xiao}}, \bibinfo {author} {\bibfnamefont {J.}~\bibnamefont {Zhao}}, \bibinfo {author} {\bibfnamefont {Y.}~\bibnamefont {Li}}, \bibinfo {author} {\bibfnamefont {R.}~\bibnamefont {Shindou}},\ and\ \bibinfo {author} {\bibfnamefont {Z.-D.}\ \bibnamefont {Song}},\ }\bibfield  {title} {\bibinfo {title} {Spin space groups: Full classification and applications},\ }\href {https://doi.org/10.1103/PhysRevX.14.031037} {\bibfield  {journal} {\bibinfo  {journal} {Phys. Rev. X}\ }\textbf {\bibinfo {volume} {14}},\ \bibinfo {pages} {031037} (\bibinfo {year} {2024})}\BibitemShut {NoStop}%
\bibitem [{\citenamefont {McClarty}\ and\ \citenamefont {Rau}(2024)}]{mcclartyLandauTheoryAltermagnetism2024}%
  \BibitemOpen
  \bibfield  {author} {\bibinfo {author} {\bibfnamefont {P.~A.}\ \bibnamefont {McClarty}}\ and\ \bibinfo {author} {\bibfnamefont {J.~G.}\ \bibnamefont {Rau}},\ }\bibfield  {title} {\bibinfo {title} {Landau theory of altermagnetism},\ }\href {https://doi.org/10.1103/PhysRevLett.132.176702} {\bibfield  {journal} {\bibinfo  {journal} {Phys. Rev. Lett.}\ }\textbf {\bibinfo {volume} {132}},\ \bibinfo {pages} {176702} (\bibinfo {year} {2024})}\BibitemShut {NoStop}%
\bibitem [{\citenamefont {Duan}\ \emph {et~al.}(2025)\citenamefont {Duan}, \citenamefont {Zhang}, \citenamefont {Zhu}, \citenamefont {Liu}, \citenamefont {Zhang}, \citenamefont {Žutić},\ and\ \citenamefont {Zhou}}]{tongzhou_prl2025a}%
  \BibitemOpen
  \bibfield  {author} {\bibinfo {author} {\bibfnamefont {X.}~\bibnamefont {Duan}}, \bibinfo {author} {\bibfnamefont {J.}~\bibnamefont {Zhang}}, \bibinfo {author} {\bibfnamefont {Z.}~\bibnamefont {Zhu}}, \bibinfo {author} {\bibfnamefont {Y.}~\bibnamefont {Liu}}, \bibinfo {author} {\bibfnamefont {Z.}~\bibnamefont {Zhang}}, \bibinfo {author} {\bibfnamefont {I.}~\bibnamefont {Žutić}},\ and\ \bibinfo {author} {\bibfnamefont {T.}~\bibnamefont {Zhou}},\ }\bibfield  {title} {\bibinfo {title} {Antiferroelectric {{Altermagnets}}: {{Antiferroelectricity Alters Magnets}}},\ }\href {https://doi.org/10.1103/PhysRevLett.134.106801} {\bibfield  {journal} {\bibinfo  {journal} {Phys. Rev. Lett.}\ }\textbf {\bibinfo {volume} {134}},\ \bibinfo {pages} {106801} (\bibinfo {year} {2025})}\BibitemShut {NoStop}%
\bibitem [{\citenamefont {Zhu}\ \emph {et~al.}(2025{\natexlab{a}})\citenamefont {Zhu}, \citenamefont {Duan}, \citenamefont {Zhang}, \citenamefont {Hao}, \citenamefont {{\v Z}uti{\'c}},\ and\ \citenamefont {Zhou}}]{zhu2025twodimensional}%
  \BibitemOpen
  \bibfield  {author} {\bibinfo {author} {\bibfnamefont {Z.}~\bibnamefont {Zhu}}, \bibinfo {author} {\bibfnamefont {X.}~\bibnamefont {Duan}}, \bibinfo {author} {\bibfnamefont {J.}~\bibnamefont {Zhang}}, \bibinfo {author} {\bibfnamefont {B.}~\bibnamefont {Hao}}, \bibinfo {author} {\bibfnamefont {I.}~\bibnamefont {{\v Z}uti{\'c}}},\ and\ \bibinfo {author} {\bibfnamefont {T.}~\bibnamefont {Zhou}},\ }\bibfield  {title} {\bibinfo {title} {Two-{{Dimensional Ferroelectric Altermagnets}}: {{From Model}} to {{Material Realization}}},\ }\href {https://doi.org/10.1021/acs.nanolett.5c02121} {\bibfield  {journal} {\bibinfo  {journal} {Nano Lett.}\ }\textbf {\bibinfo {volume} {25}},\ \bibinfo {pages} {9456} (\bibinfo {year} {2025}{\natexlab{a}})}\BibitemShut {NoStop}%
\bibitem [{\citenamefont {Song}\ \emph {et~al.}(2025{\natexlab{b}})\citenamefont {Song}, \citenamefont {Bai}, \citenamefont {Zhou}, \citenamefont {Han}, \citenamefont {Reichlova}, \citenamefont {Dil}, \citenamefont {Liu}, \citenamefont {Chen},\ and\ \citenamefont {Pan}}]{song2025altermagnets}%
  \BibitemOpen
  \bibfield  {author} {\bibinfo {author} {\bibfnamefont {C.}~\bibnamefont {Song}}, \bibinfo {author} {\bibfnamefont {H.}~\bibnamefont {Bai}}, \bibinfo {author} {\bibfnamefont {Z.}~\bibnamefont {Zhou}}, \bibinfo {author} {\bibfnamefont {L.}~\bibnamefont {Han}}, \bibinfo {author} {\bibfnamefont {H.}~\bibnamefont {Reichlova}}, \bibinfo {author} {\bibfnamefont {J.~H.}\ \bibnamefont {Dil}}, \bibinfo {author} {\bibfnamefont {J.}~\bibnamefont {Liu}}, \bibinfo {author} {\bibfnamefont {X.}~\bibnamefont {Chen}},\ and\ \bibinfo {author} {\bibfnamefont {F.}~\bibnamefont {Pan}},\ }\bibfield  {title} {\bibinfo {title} {Altermagnets as a new class of functional materials},\ }\href {https://doi.org/10.1038/s41578-025-00779-1} {\bibfield  {journal} {\bibinfo  {journal} {Nat. Rev. Mater.}\ }\textbf {\bibinfo {volume} {10}},\ \bibinfo {pages} {473} (\bibinfo {year} {2025}{\natexlab{b}})}\BibitemShut {NoStop}%
\bibitem [{\citenamefont {Zhang}\ \emph {et~al.}(2025)\citenamefont {Zhang}, \citenamefont {Cheng}, \citenamefont {Yin}, \citenamefont {Liu}, \citenamefont {Deng}, \citenamefont {Qiao}, \citenamefont {Shi}, \citenamefont {Zhang}, \citenamefont {Lin}, \citenamefont {Liu}, \citenamefont {Ye}, \citenamefont {Huang}, \citenamefont {Meng}, \citenamefont {Zhang}, \citenamefont {Okuda}, \citenamefont {Shimada}, \citenamefont {Cui}, \citenamefont {Zhao}, \citenamefont {Cao}, \citenamefont {Qiao}, \citenamefont {Liu},\ and\ \citenamefont {Chen}}]{zhang2025crystalsymmetrypaired}%
  \BibitemOpen
  \bibfield  {author} {\bibinfo {author} {\bibfnamefont {F.}~\bibnamefont {Zhang}}, \bibinfo {author} {\bibfnamefont {X.}~\bibnamefont {Cheng}}, \bibinfo {author} {\bibfnamefont {Z.}~\bibnamefont {Yin}}, \bibinfo {author} {\bibfnamefont {C.}~\bibnamefont {Liu}}, \bibinfo {author} {\bibfnamefont {L.}~\bibnamefont {Deng}}, \bibinfo {author} {\bibfnamefont {Y.}~\bibnamefont {Qiao}}, \bibinfo {author} {\bibfnamefont {Z.}~\bibnamefont {Shi}}, \bibinfo {author} {\bibfnamefont {S.}~\bibnamefont {Zhang}}, \bibinfo {author} {\bibfnamefont {J.}~\bibnamefont {Lin}}, \bibinfo {author} {\bibfnamefont {Z.}~\bibnamefont {Liu}}, \bibinfo {author} {\bibfnamefont {M.}~\bibnamefont {Ye}}, \bibinfo {author} {\bibfnamefont {Y.}~\bibnamefont {Huang}}, \bibinfo {author} {\bibfnamefont {X.}~\bibnamefont {Meng}}, \bibinfo {author} {\bibfnamefont {C.}~\bibnamefont {Zhang}}, \bibinfo {author} {\bibfnamefont {T.}~\bibnamefont {Okuda}}, \bibinfo {author} {\bibfnamefont {K.}~\bibnamefont {Shimada}}, \bibinfo {author} {\bibfnamefont
  {S.}~\bibnamefont {Cui}}, \bibinfo {author} {\bibfnamefont {Y.}~\bibnamefont {Zhao}}, \bibinfo {author} {\bibfnamefont {G.-H.}\ \bibnamefont {Cao}}, \bibinfo {author} {\bibfnamefont {S.}~\bibnamefont {Qiao}}, \bibinfo {author} {\bibfnamefont {J.}~\bibnamefont {Liu}},\ and\ \bibinfo {author} {\bibfnamefont {C.}~\bibnamefont {Chen}},\ }\bibfield  {title} {\bibinfo {title} {Crystal-symmetry-paired spin--valley locking in a layered room-temperature metallic altermagnet candidate},\ }\href {https://doi.org/10.1038/s41567-025-02864-2} {\bibfield  {journal} {\bibinfo  {journal} {Nat. Phys.}\ }\textbf {\bibinfo {volume} {21}},\ \bibinfo {pages} {760} (\bibinfo {year} {2025})}\BibitemShut {NoStop}%
\bibitem [{\citenamefont {Amin}\ \emph {et~al.}(2024)\citenamefont {Amin}, \citenamefont {Dal~Din}, \citenamefont {Golias}, \citenamefont {Niu}, \citenamefont {Zakharov}, \citenamefont {Fromage}, \citenamefont {Fields}, \citenamefont {Heywood}, \citenamefont {Cousins}, \citenamefont {Maccherozzi}, \citenamefont {Krempask{\'y}}, \citenamefont {Dil}, \citenamefont {Kriegner}, \citenamefont {Kiraly}, \citenamefont {Campion}, \citenamefont {Rushforth}, \citenamefont {Edmonds}, \citenamefont {Dhesi}, \citenamefont {{\v S}mejkal}, \citenamefont {Jungwirth},\ and\ \citenamefont {Wadley}}]{aminNanoscaleImagingControl2024}%
  \BibitemOpen
  \bibfield  {author} {\bibinfo {author} {\bibfnamefont {O.~J.}\ \bibnamefont {Amin}}, \bibinfo {author} {\bibfnamefont {A.}~\bibnamefont {Dal~Din}}, \bibinfo {author} {\bibfnamefont {E.}~\bibnamefont {Golias}}, \bibinfo {author} {\bibfnamefont {Y.}~\bibnamefont {Niu}}, \bibinfo {author} {\bibfnamefont {A.}~\bibnamefont {Zakharov}}, \bibinfo {author} {\bibfnamefont {S.~C.}\ \bibnamefont {Fromage}}, \bibinfo {author} {\bibfnamefont {C.~J.~B.}\ \bibnamefont {Fields}}, \bibinfo {author} {\bibfnamefont {S.~L.}\ \bibnamefont {Heywood}}, \bibinfo {author} {\bibfnamefont {R.~B.}\ \bibnamefont {Cousins}}, \bibinfo {author} {\bibfnamefont {F.}~\bibnamefont {Maccherozzi}}, \bibinfo {author} {\bibfnamefont {J.}~\bibnamefont {Krempask{\'y}}}, \bibinfo {author} {\bibfnamefont {J.~H.}\ \bibnamefont {Dil}}, \bibinfo {author} {\bibfnamefont {D.}~\bibnamefont {Kriegner}}, \bibinfo {author} {\bibfnamefont {B.}~\bibnamefont {Kiraly}}, \bibinfo {author} {\bibfnamefont {R.~P.}\ \bibnamefont {Campion}}, \bibinfo {author}
  {\bibfnamefont {A.~W.}\ \bibnamefont {Rushforth}}, \bibinfo {author} {\bibfnamefont {K.~W.}\ \bibnamefont {Edmonds}}, \bibinfo {author} {\bibfnamefont {S.~S.}\ \bibnamefont {Dhesi}}, \bibinfo {author} {\bibfnamefont {L.}~\bibnamefont {{\v S}mejkal}}, \bibinfo {author} {\bibfnamefont {T.}~\bibnamefont {Jungwirth}},\ and\ \bibinfo {author} {\bibfnamefont {P.}~\bibnamefont {Wadley}},\ }\bibfield  {title} {\bibinfo {title} {Nanoscale imaging and control of altermagnetism in mnte},\ }\href {https://doi.org/10.1038/s41586-024-08234-x} {\bibfield  {journal} {\bibinfo  {journal} {Nature}\ }\textbf {\bibinfo {volume} {636}},\ \bibinfo {pages} {348} (\bibinfo {year} {2024})}\BibitemShut {NoStop}%
\bibitem [{\citenamefont {Gonzalez~Betancourt}\ \emph {et~al.}(2024)\citenamefont {Gonzalez~Betancourt}, \citenamefont {Zub{\'a}{\v c}}, \citenamefont {Geishendorf}, \citenamefont {Ritzinger}, \citenamefont {R{\r u}{\v z}i{\v c}kov{\'a}}, \citenamefont {Kotte}, \citenamefont {{\v Z}elezn{\'y}}, \citenamefont {Olejn{\'i}k}, \citenamefont {Springholz}, \citenamefont {B{\"u}chner}, \citenamefont {Thomas}, \citenamefont {V{\'y}born{\'y}}, \citenamefont {Jungwirth}, \citenamefont {Reichlov{\'a}},\ and\ \citenamefont {Kriegner}}]{gonzalezbetancourt2024anisotropic}%
  \BibitemOpen
  \bibfield  {author} {\bibinfo {author} {\bibfnamefont {R.~D.}\ \bibnamefont {Gonzalez~Betancourt}}, \bibinfo {author} {\bibfnamefont {J.}~\bibnamefont {Zub{\'a}{\v c}}}, \bibinfo {author} {\bibfnamefont {K.}~\bibnamefont {Geishendorf}}, \bibinfo {author} {\bibfnamefont {P.}~\bibnamefont {Ritzinger}}, \bibinfo {author} {\bibfnamefont {B.}~\bibnamefont {R{\r u}{\v z}i{\v c}kov{\'a}}}, \bibinfo {author} {\bibfnamefont {T.}~\bibnamefont {Kotte}}, \bibinfo {author} {\bibfnamefont {J.}~\bibnamefont {{\v Z}elezn{\'y}}}, \bibinfo {author} {\bibfnamefont {K.}~\bibnamefont {Olejn{\'i}k}}, \bibinfo {author} {\bibfnamefont {G.}~\bibnamefont {Springholz}}, \bibinfo {author} {\bibfnamefont {B.}~\bibnamefont {B{\"u}chner}}, \bibinfo {author} {\bibfnamefont {A.}~\bibnamefont {Thomas}}, \bibinfo {author} {\bibfnamefont {K.}~\bibnamefont {V{\'y}born{\'y}}}, \bibinfo {author} {\bibfnamefont {T.}~\bibnamefont {Jungwirth}}, \bibinfo {author} {\bibfnamefont {H.}~\bibnamefont {Reichlov{\'a}}},\ and\ \bibinfo {author}
  {\bibfnamefont {D.}~\bibnamefont {Kriegner}},\ }\bibfield  {title} {\bibinfo {title} {Anisotropic magnetoresistance in altermagnetic mnte},\ }\href {https://doi.org/10.1038/s44306-024-00046-z} {\bibfield  {journal} {\bibinfo  {journal} {npj Spintronics}\ }\textbf {\bibinfo {volume} {2}},\ \bibinfo {pages} {45} (\bibinfo {year} {2024})}\BibitemShut {NoStop}%
\bibitem [{\citenamefont {Lee}\ \emph {et~al.}(2024)\citenamefont {Lee}, \citenamefont {Lee}, \citenamefont {Jung}, \citenamefont {Jung}, \citenamefont {Kim}, \citenamefont {Lee}, \citenamefont {Seok}, \citenamefont {Kim}, \citenamefont {Park}, \citenamefont {{\v S}mejkal}, \citenamefont {Kang},\ and\ \citenamefont {Kim}}]{PhysRevLett.132.036702}%
  \BibitemOpen
  \bibfield  {author} {\bibinfo {author} {\bibfnamefont {S.}~\bibnamefont {Lee}}, \bibinfo {author} {\bibfnamefont {S.}~\bibnamefont {Lee}}, \bibinfo {author} {\bibfnamefont {S.}~\bibnamefont {Jung}}, \bibinfo {author} {\bibfnamefont {J.}~\bibnamefont {Jung}}, \bibinfo {author} {\bibfnamefont {D.}~\bibnamefont {Kim}}, \bibinfo {author} {\bibfnamefont {Y.}~\bibnamefont {Lee}}, \bibinfo {author} {\bibfnamefont {B.}~\bibnamefont {Seok}}, \bibinfo {author} {\bibfnamefont {J.}~\bibnamefont {Kim}}, \bibinfo {author} {\bibfnamefont {B.~G.}\ \bibnamefont {Park}}, \bibinfo {author} {\bibfnamefont {L.}~\bibnamefont {{\v S}mejkal}}, \bibinfo {author} {\bibfnamefont {C.-J.}\ \bibnamefont {Kang}},\ and\ \bibinfo {author} {\bibfnamefont {C.}~\bibnamefont {Kim}},\ }\bibfield  {title} {\bibinfo {title} {Broken kramers degeneracy in altermagnetic mnte},\ }\href {https://doi.org/10.1103/PhysRevLett.132.036702} {\bibfield  {journal} {\bibinfo  {journal} {Phys. Rev. Lett.}\ }\textbf {\bibinfo {volume} {132}},\ \bibinfo {pages}
  {036702} (\bibinfo {year} {2024})}\BibitemShut {NoStop}%
\bibitem [{\citenamefont {Reimers}\ \emph {et~al.}(2024)\citenamefont {Reimers}, \citenamefont {Odenbreit}, \citenamefont {{\v S}mejkal}, \citenamefont {Strocov}, \citenamefont {Constantinou}, \citenamefont {Hellenes}, \citenamefont {Jaeschke~Ubiergo}, \citenamefont {Campos}, \citenamefont {Bharadwaj}, \citenamefont {Chakraborty}, \citenamefont {Denneulin}, \citenamefont {Shi}, \citenamefont {{Dunin-Borkowski}}, \citenamefont {Das}, \citenamefont {Kl{\"a}ui}, \citenamefont {Sinova},\ and\ \citenamefont {Jourdan}}]{reimersDirectObservationAltermagnetic2024a}%
  \BibitemOpen
  \bibfield  {author} {\bibinfo {author} {\bibfnamefont {S.}~\bibnamefont {Reimers}}, \bibinfo {author} {\bibfnamefont {L.}~\bibnamefont {Odenbreit}}, \bibinfo {author} {\bibfnamefont {L.}~\bibnamefont {{\v S}mejkal}}, \bibinfo {author} {\bibfnamefont {V.~N.}\ \bibnamefont {Strocov}}, \bibinfo {author} {\bibfnamefont {P.}~\bibnamefont {Constantinou}}, \bibinfo {author} {\bibfnamefont {A.~B.}\ \bibnamefont {Hellenes}}, \bibinfo {author} {\bibfnamefont {R.}~\bibnamefont {Jaeschke~Ubiergo}}, \bibinfo {author} {\bibfnamefont {W.~H.}\ \bibnamefont {Campos}}, \bibinfo {author} {\bibfnamefont {V.~K.}\ \bibnamefont {Bharadwaj}}, \bibinfo {author} {\bibfnamefont {A.}~\bibnamefont {Chakraborty}}, \bibinfo {author} {\bibfnamefont {T.}~\bibnamefont {Denneulin}}, \bibinfo {author} {\bibfnamefont {W.}~\bibnamefont {Shi}}, \bibinfo {author} {\bibfnamefont {R.~E.}\ \bibnamefont {{Dunin-Borkowski}}}, \bibinfo {author} {\bibfnamefont {S.}~\bibnamefont {Das}}, \bibinfo {author} {\bibfnamefont {M.}~\bibnamefont {Kl{\"a}ui}},
  \bibinfo {author} {\bibfnamefont {J.}~\bibnamefont {Sinova}},\ and\ \bibinfo {author} {\bibfnamefont {M.}~\bibnamefont {Jourdan}},\ }\bibfield  {title} {\bibinfo {title} {Direct observation of altermagnetic band splitting in crsb thin films},\ }\href {https://doi.org/10.1038/s41467-024-46476-5} {\bibfield  {journal} {\bibinfo  {journal} {Nat. Commun.}\ }\textbf {\bibinfo {volume} {15}},\ \bibinfo {pages} {2116} (\bibinfo {year} {2024})}\BibitemShut {NoStop}%
\bibitem [{\citenamefont {Zhou}\ \emph {et~al.}(2025)\citenamefont {Zhou}, \citenamefont {Cheng}, \citenamefont {Hu}, \citenamefont {Chu}, \citenamefont {Bai}, \citenamefont {Han}, \citenamefont {Liu}, \citenamefont {Pan},\ and\ \citenamefont {Song}}]{zhouManipulationAltermagneticOrder2025}%
  \BibitemOpen
  \bibfield  {author} {\bibinfo {author} {\bibfnamefont {Z.}~\bibnamefont {Zhou}}, \bibinfo {author} {\bibfnamefont {X.}~\bibnamefont {Cheng}}, \bibinfo {author} {\bibfnamefont {M.}~\bibnamefont {Hu}}, \bibinfo {author} {\bibfnamefont {R.}~\bibnamefont {Chu}}, \bibinfo {author} {\bibfnamefont {H.}~\bibnamefont {Bai}}, \bibinfo {author} {\bibfnamefont {L.}~\bibnamefont {Han}}, \bibinfo {author} {\bibfnamefont {J.}~\bibnamefont {Liu}}, \bibinfo {author} {\bibfnamefont {F.}~\bibnamefont {Pan}},\ and\ \bibinfo {author} {\bibfnamefont {C.}~\bibnamefont {Song}},\ }\bibfield  {title} {\bibinfo {title} {Manipulation of the altermagnetic order in crsb via crystal symmetry},\ }\href {https://doi.org/10.1038/s41586-024-08436-3} {\bibfield  {journal} {\bibinfo  {journal} {Nature}\ }\textbf {\bibinfo {volume} {638}},\ \bibinfo {pages} {645} (\bibinfo {year} {2025})}\BibitemShut {NoStop}%
\bibitem [{\citenamefont {Yang}\ \emph {et~al.}(2025)\citenamefont {Yang}, \citenamefont {Li}, \citenamefont {Yang}, \citenamefont {Li}, \citenamefont {Zheng}, \citenamefont {Zhu}, \citenamefont {Pan}, \citenamefont {Xu}, \citenamefont {Cao}, \citenamefont {Zhao}, \citenamefont {Jana}, \citenamefont {Zhang}, \citenamefont {Ye}, \citenamefont {Song}, \citenamefont {Hu}, \citenamefont {Yang}, \citenamefont {Fujii}, \citenamefont {Vobornik}, \citenamefont {Shi}, \citenamefont {Yuan}, \citenamefont {Zhang}, \citenamefont {Xu},\ and\ \citenamefont {Liu}}]{yang2025threedimensional}%
  \BibitemOpen
  \bibfield  {author} {\bibinfo {author} {\bibfnamefont {G.}~\bibnamefont {Yang}}, \bibinfo {author} {\bibfnamefont {Z.}~\bibnamefont {Li}}, \bibinfo {author} {\bibfnamefont {S.}~\bibnamefont {Yang}}, \bibinfo {author} {\bibfnamefont {J.}~\bibnamefont {Li}}, \bibinfo {author} {\bibfnamefont {H.}~\bibnamefont {Zheng}}, \bibinfo {author} {\bibfnamefont {W.}~\bibnamefont {Zhu}}, \bibinfo {author} {\bibfnamefont {Z.}~\bibnamefont {Pan}}, \bibinfo {author} {\bibfnamefont {Y.}~\bibnamefont {Xu}}, \bibinfo {author} {\bibfnamefont {S.}~\bibnamefont {Cao}}, \bibinfo {author} {\bibfnamefont {W.}~\bibnamefont {Zhao}}, \bibinfo {author} {\bibfnamefont {A.}~\bibnamefont {Jana}}, \bibinfo {author} {\bibfnamefont {J.}~\bibnamefont {Zhang}}, \bibinfo {author} {\bibfnamefont {M.}~\bibnamefont {Ye}}, \bibinfo {author} {\bibfnamefont {Y.}~\bibnamefont {Song}}, \bibinfo {author} {\bibfnamefont {L.-H.}\ \bibnamefont {Hu}}, \bibinfo {author} {\bibfnamefont {L.}~\bibnamefont {Yang}}, \bibinfo {author} {\bibfnamefont
  {J.}~\bibnamefont {Fujii}}, \bibinfo {author} {\bibfnamefont {I.}~\bibnamefont {Vobornik}}, \bibinfo {author} {\bibfnamefont {M.}~\bibnamefont {Shi}}, \bibinfo {author} {\bibfnamefont {H.}~\bibnamefont {Yuan}}, \bibinfo {author} {\bibfnamefont {Y.}~\bibnamefont {Zhang}}, \bibinfo {author} {\bibfnamefont {Y.}~\bibnamefont {Xu}},\ and\ \bibinfo {author} {\bibfnamefont {Y.}~\bibnamefont {Liu}},\ }\bibfield  {title} {\bibinfo {title} {Three-dimensional mapping of the altermagnetic spin splitting in crsb},\ }\href {https://doi.org/10.1038/s41467-025-56647-7} {\bibfield  {journal} {\bibinfo  {journal} {Nat. Commun.}\ }\textbf {\bibinfo {volume} {16}},\ \bibinfo {pages} {1442} (\bibinfo {year} {2025})}\BibitemShut {NoStop}%
\bibitem [{\citenamefont {Leivisk{\"a}}\ \emph {et~al.}(2024)\citenamefont {Leivisk{\"a}}, \citenamefont {Rial}, \citenamefont {Bad'ura}, \citenamefont {Seeger}, \citenamefont {Kounta}, \citenamefont {Beckert}, \citenamefont {Kriegner}, \citenamefont {Joumard}, \citenamefont {Schmoranzerov{\'a}}, \citenamefont {Sinova}, \citenamefont {Gomonay}, \citenamefont {Thomas}, \citenamefont {Goennenwein}, \citenamefont {Reichlov{\'a}}, \citenamefont {{\v S}~{\v S}mejkal}, \citenamefont {Michez}, \citenamefont {{\v s}~Jungwirth},\ and\ \citenamefont {Baltz}}]{PhysRevB.109.224430}%
  \BibitemOpen
  \bibfield  {author} {\bibinfo {author} {\bibfnamefont {M.}~\bibnamefont {Leivisk{\"a}}}, \bibinfo {author} {\bibfnamefont {J.}~\bibnamefont {Rial}}, \bibinfo {author} {\bibfnamefont {A.}~\bibnamefont {Bad'ura}}, \bibinfo {author} {\bibfnamefont {R.~L.}\ \bibnamefont {Seeger}}, \bibinfo {author} {\bibfnamefont {I.}~\bibnamefont {Kounta}}, \bibinfo {author} {\bibfnamefont {S.}~\bibnamefont {Beckert}}, \bibinfo {author} {\bibfnamefont {D.}~\bibnamefont {Kriegner}}, \bibinfo {author} {\bibfnamefont {I.}~\bibnamefont {Joumard}}, \bibinfo {author} {\bibfnamefont {E.}~\bibnamefont {Schmoranzerov{\'a}}}, \bibinfo {author} {\bibfnamefont {J.}~\bibnamefont {Sinova}}, \bibinfo {author} {\bibfnamefont {O.}~\bibnamefont {Gomonay}}, \bibinfo {author} {\bibfnamefont {A.}~\bibnamefont {Thomas}}, \bibinfo {author} {\bibfnamefont {S.~T.~B.}\ \bibnamefont {Goennenwein}}, \bibinfo {author} {\bibfnamefont {H.}~\bibnamefont {Reichlov{\'a}}}, \bibinfo {author} {\bibfnamefont {L.}~\bibnamefont {{\v S}~{\v S}mejkal}}, \bibinfo
  {author} {\bibfnamefont {L.}~\bibnamefont {Michez}}, \bibinfo {author} {\bibfnamefont {T.}~\bibnamefont {{\v s}~Jungwirth}},\ and\ \bibinfo {author} {\bibfnamefont {V.}~\bibnamefont {Baltz}},\ }\bibfield  {title} {\bibinfo {title} {Anisotropy of the anomalous hall effect in thin films of the altermagnet candidate {Mn{$_5$}Si{$_{3}$}}},\ }\href {https://doi.org/10.1103/PhysRevB.109.224430} {\bibfield  {journal} {\bibinfo  {journal} {Phys. Rev. B}\ }\textbf {\bibinfo {volume} {109}},\ \bibinfo {pages} {224430} (\bibinfo {year} {2024})}\BibitemShut {NoStop}%
\bibitem [{\citenamefont {Badura}\ \emph {et~al.}(2025)\citenamefont {Badura}, \citenamefont {Campos}, \citenamefont {Bharadwaj}, \citenamefont {Kounta}, \citenamefont {Michez}, \citenamefont {Petit}, \citenamefont {Rial}, \citenamefont {Leivisk{\"a}}, \citenamefont {Baltz}, \citenamefont {Krizek}, \citenamefont {Kriegner}, \citenamefont {{\v Z}elezn{\'y}}, \citenamefont {Zemen}, \citenamefont {Telkamp}, \citenamefont {Sailler}, \citenamefont {Lammel}, \citenamefont {{Jaeschke-Ubiergo}}, \citenamefont {Hellenes}, \citenamefont {{Gonz{\'a}lez-Hern{\'a}ndez}}, \citenamefont {Sinova}, \citenamefont {Jungwirth}, \citenamefont {Goennenwein}, \citenamefont {{\v S}mejkal},\ and\ \citenamefont {Reichlova}}]{badura2025observation}%
  \BibitemOpen
  \bibfield  {author} {\bibinfo {author} {\bibfnamefont {A.}~\bibnamefont {Badura}}, \bibinfo {author} {\bibfnamefont {W.~H.}\ \bibnamefont {Campos}}, \bibinfo {author} {\bibfnamefont {V.~K.}\ \bibnamefont {Bharadwaj}}, \bibinfo {author} {\bibfnamefont {I.}~\bibnamefont {Kounta}}, \bibinfo {author} {\bibfnamefont {L.}~\bibnamefont {Michez}}, \bibinfo {author} {\bibfnamefont {M.}~\bibnamefont {Petit}}, \bibinfo {author} {\bibfnamefont {J.}~\bibnamefont {Rial}}, \bibinfo {author} {\bibfnamefont {M.}~\bibnamefont {Leivisk{\"a}}}, \bibinfo {author} {\bibfnamefont {V.}~\bibnamefont {Baltz}}, \bibinfo {author} {\bibfnamefont {F.}~\bibnamefont {Krizek}}, \bibinfo {author} {\bibfnamefont {D.}~\bibnamefont {Kriegner}}, \bibinfo {author} {\bibfnamefont {J.}~\bibnamefont {{\v Z}elezn{\'y}}}, \bibinfo {author} {\bibfnamefont {J.}~\bibnamefont {Zemen}}, \bibinfo {author} {\bibfnamefont {S.}~\bibnamefont {Telkamp}}, \bibinfo {author} {\bibfnamefont {S.}~\bibnamefont {Sailler}}, \bibinfo {author} {\bibfnamefont
  {M.}~\bibnamefont {Lammel}}, \bibinfo {author} {\bibfnamefont {R.}~\bibnamefont {{Jaeschke-Ubiergo}}}, \bibinfo {author} {\bibfnamefont {A.~B.}\ \bibnamefont {Hellenes}}, \bibinfo {author} {\bibfnamefont {R.}~\bibnamefont {{Gonz{\'a}lez-Hern{\'a}ndez}}}, \bibinfo {author} {\bibfnamefont {J.}~\bibnamefont {Sinova}}, \bibinfo {author} {\bibfnamefont {T.}~\bibnamefont {Jungwirth}}, \bibinfo {author} {\bibfnamefont {S.~T.~B.}\ \bibnamefont {Goennenwein}}, \bibinfo {author} {\bibfnamefont {L.}~\bibnamefont {{\v S}mejkal}},\ and\ \bibinfo {author} {\bibfnamefont {H.}~\bibnamefont {Reichlova}},\ }\bibfield  {title} {\bibinfo {title} {Observation of the anomalous nernst effect in altermagnetic candidate {Mn{$_5$}Si{$_{3}$}}},\ }\href {https://doi.org/10.1038/s41467-025-62331-7} {\bibfield  {journal} {\bibinfo  {journal} {Nat. Commun.}\ }\textbf {\bibinfo {volume} {16}},\ \bibinfo {pages} {7111} (\bibinfo {year} {2025})}\BibitemShut {NoStop}%
\bibitem [{\citenamefont {Jiang}\ \emph {et~al.}(2025)\citenamefont {Jiang}, \citenamefont {Hu}, \citenamefont {Bai}, \citenamefont {Song}, \citenamefont {Mu}, \citenamefont {Qu}, \citenamefont {Li}, \citenamefont {Zhu}, \citenamefont {Pi}, \citenamefont {Wei}, \citenamefont {Sun}, \citenamefont {Huang}, \citenamefont {Zheng}, \citenamefont {Peng}, \citenamefont {He}, \citenamefont {Li}, \citenamefont {Luo}, \citenamefont {Li}, \citenamefont {Chen}, \citenamefont {Li}, \citenamefont {Weng},\ and\ \citenamefont {Qian}}]{jiang2025metallic}%
  \BibitemOpen
  \bibfield  {author} {\bibinfo {author} {\bibfnamefont {B.}~\bibnamefont {Jiang}}, \bibinfo {author} {\bibfnamefont {M.}~\bibnamefont {Hu}}, \bibinfo {author} {\bibfnamefont {J.}~\bibnamefont {Bai}}, \bibinfo {author} {\bibfnamefont {Z.}~\bibnamefont {Song}}, \bibinfo {author} {\bibfnamefont {C.}~\bibnamefont {Mu}}, \bibinfo {author} {\bibfnamefont {G.}~\bibnamefont {Qu}}, \bibinfo {author} {\bibfnamefont {W.}~\bibnamefont {Li}}, \bibinfo {author} {\bibfnamefont {W.}~\bibnamefont {Zhu}}, \bibinfo {author} {\bibfnamefont {H.}~\bibnamefont {Pi}}, \bibinfo {author} {\bibfnamefont {Z.}~\bibnamefont {Wei}}, \bibinfo {author} {\bibfnamefont {Y.-J.}\ \bibnamefont {Sun}}, \bibinfo {author} {\bibfnamefont {Y.}~\bibnamefont {Huang}}, \bibinfo {author} {\bibfnamefont {X.}~\bibnamefont {Zheng}}, \bibinfo {author} {\bibfnamefont {Y.}~\bibnamefont {Peng}}, \bibinfo {author} {\bibfnamefont {L.}~\bibnamefont {He}}, \bibinfo {author} {\bibfnamefont {S.}~\bibnamefont {Li}}, \bibinfo {author} {\bibfnamefont {J.}~\bibnamefont
  {Luo}}, \bibinfo {author} {\bibfnamefont {Z.}~\bibnamefont {Li}}, \bibinfo {author} {\bibfnamefont {G.}~\bibnamefont {Chen}}, \bibinfo {author} {\bibfnamefont {H.}~\bibnamefont {Li}}, \bibinfo {author} {\bibfnamefont {H.}~\bibnamefont {Weng}},\ and\ \bibinfo {author} {\bibfnamefont {T.}~\bibnamefont {Qian}},\ }\bibfield  {title} {\bibinfo {title} {A metallic room-temperature d-wave altermagnet},\ }\href {https://doi.org/10.1038/s41567-025-02822-y} {\bibfield  {journal} {\bibinfo  {journal} {Nat. Phys.}\ }\textbf {\bibinfo {volume} {21}},\ \bibinfo {pages} {754} (\bibinfo {year} {2025})}\BibitemShut {NoStop}%
\bibitem [{\citenamefont {Wang}\ \emph {et~al.}(2025{\natexlab{a}})\citenamefont {Wang}, \citenamefont {Yu}, \citenamefont {Cheng}, \citenamefont {Xiao}, \citenamefont {Ma}, \citenamefont {Quan}, \citenamefont {Song}, \citenamefont {Zhang}, \citenamefont {Zhang}, \citenamefont {Ma}, \citenamefont {Liu}, \citenamefont {Yadav}, \citenamefont {Shi}, \citenamefont {Wang}, \citenamefont {Niu}, \citenamefont {Gao}, \citenamefont {Xiang}, \citenamefont {Liu}, \citenamefont {Wang},\ and\ \citenamefont {Chen}}]{wang2025atomicscalespinsensing2d}%
  \BibitemOpen
  \bibfield  {author} {\bibinfo {author} {\bibfnamefont {Z.}~\bibnamefont {Wang}}, \bibinfo {author} {\bibfnamefont {S.}~\bibnamefont {Yu}}, \bibinfo {author} {\bibfnamefont {X.}~\bibnamefont {Cheng}}, \bibinfo {author} {\bibfnamefont {X.}~\bibnamefont {Xiao}}, \bibinfo {author} {\bibfnamefont {W.}~\bibnamefont {Ma}}, \bibinfo {author} {\bibfnamefont {F.}~\bibnamefont {Quan}}, \bibinfo {author} {\bibfnamefont {H.}~\bibnamefont {Song}}, \bibinfo {author} {\bibfnamefont {K.}~\bibnamefont {Zhang}}, \bibinfo {author} {\bibfnamefont {Y.}~\bibnamefont {Zhang}}, \bibinfo {author} {\bibfnamefont {Y.}~\bibnamefont {Ma}}, \bibinfo {author} {\bibfnamefont {W.}~\bibnamefont {Liu}}, \bibinfo {author} {\bibfnamefont {P.}~\bibnamefont {Yadav}}, \bibinfo {author} {\bibfnamefont {X.}~\bibnamefont {Shi}}, \bibinfo {author} {\bibfnamefont {Z.}~\bibnamefont {Wang}}, \bibinfo {author} {\bibfnamefont {Q.}~\bibnamefont {Niu}}, \bibinfo {author} {\bibfnamefont {Y.}~\bibnamefont {Gao}}, \bibinfo {author} {\bibfnamefont
  {B.}~\bibnamefont {Xiang}}, \bibinfo {author} {\bibfnamefont {J.}~\bibnamefont {Liu}}, \bibinfo {author} {\bibfnamefont {Z.}~\bibnamefont {Wang}},\ and\ \bibinfo {author} {\bibfnamefont {X.}~\bibnamefont {Chen}},\ }\bibfield  {title} {\bibinfo {title} {Atomic-scale spin sensing of a 2d $d$-wave altermagnet via helical tunneling},\ }\href {https://arxiv.org/abs/2512.23290} {\bibfield  {journal} {\bibinfo  {journal} {arXiv:2512.23290}\ } (\bibinfo {year} {2025}{\natexlab{a}})}\BibitemShut {NoStop}%
\bibitem [{\citenamefont {{\v S}mejkal}\ \emph {et~al.}(2020)\citenamefont {{\v S}mejkal}, \citenamefont {{Gonz{\'a}lez-Hern{\'a}ndez}}, \citenamefont {Jungwirth},\ and\ \citenamefont {Sinova}}]{doi:10.1126/sciadv.aaz8809}%
  \BibitemOpen
  \bibfield  {author} {\bibinfo {author} {\bibfnamefont {L.}~\bibnamefont {{\v S}mejkal}}, \bibinfo {author} {\bibfnamefont {R.}~\bibnamefont {{Gonz{\'a}lez-Hern{\'a}ndez}}}, \bibinfo {author} {\bibfnamefont {T.}~\bibnamefont {Jungwirth}},\ and\ \bibinfo {author} {\bibfnamefont {J.}~\bibnamefont {Sinova}},\ }\bibfield  {title} {\bibinfo {title} {Crystal time-reversal symmetry breaking and spontaneous hall effect in collinear antiferromagnets},\ }\href {https://doi.org/10.1126/sciadv.aaz8809} {\bibfield  {journal} {\bibinfo  {journal} {Sci. Adv.}\ }\textbf {\bibinfo {volume} {6}},\ \bibinfo {pages} {eaaz8809} (\bibinfo {year} {2020})}\BibitemShut {NoStop}%
\bibitem [{\citenamefont {Sato}\ \emph {et~al.}(2024)\citenamefont {Sato}, \citenamefont {Haddad}, \citenamefont {Fulga}, \citenamefont {Assaad},\ and\ \citenamefont {van~den Brink}}]{sato2024altermagnetic}%
  \BibitemOpen
  \bibfield  {author} {\bibinfo {author} {\bibfnamefont {T.}~\bibnamefont {Sato}}, \bibinfo {author} {\bibfnamefont {S.}~\bibnamefont {Haddad}}, \bibinfo {author} {\bibfnamefont {I.~C.}\ \bibnamefont {Fulga}}, \bibinfo {author} {\bibfnamefont {F.~F.}\ \bibnamefont {Assaad}},\ and\ \bibinfo {author} {\bibfnamefont {J.}~\bibnamefont {van~den Brink}},\ }\bibfield  {title} {\bibinfo {title} {Altermagnetic anomalous hall effect emerging from electronic correlations},\ }\href {https://doi.org/10.1103/PhysRevLett.133.086503} {\bibfield  {journal} {\bibinfo  {journal} {Phys. Rev. Lett.}\ }\textbf {\bibinfo {volume} {133}},\ \bibinfo {pages} {086503} (\bibinfo {year} {2024})}\BibitemShut {NoStop}%
\bibitem [{\citenamefont {Brekke}\ \emph {et~al.}(2024)\citenamefont {Brekke}, \citenamefont {Sukhachov}, \citenamefont {Giil}, \citenamefont {Brataas},\ and\ \citenamefont {Linder}}]{brekkeMinimalModelsTransport2024a}%
  \BibitemOpen
  \bibfield  {author} {\bibinfo {author} {\bibfnamefont {B.}~\bibnamefont {Brekke}}, \bibinfo {author} {\bibfnamefont {P.}~\bibnamefont {Sukhachov}}, \bibinfo {author} {\bibfnamefont {H.~G.}\ \bibnamefont {Giil}}, \bibinfo {author} {\bibfnamefont {A.}~\bibnamefont {Brataas}},\ and\ \bibinfo {author} {\bibfnamefont {J.}~\bibnamefont {Linder}},\ }\bibfield  {title} {\bibinfo {title} {Minimal {{Models}} and {{Transport Properties}} of {{Unconventional}} p -{{Wave Magnets}}},\ }\href {https://doi.org/10.1103/PhysRevLett.133.236703} {\bibfield  {journal} {\bibinfo  {journal} {Phys. Rev. Lett.}\ }\textbf {\bibinfo {volume} {133}},\ \bibinfo {pages} {236703} (\bibinfo {year} {2024})}\BibitemShut {NoStop}%
\bibitem [{\citenamefont {Jin}\ \emph {et~al.}(2024)\citenamefont {Jin}, \citenamefont {Zeng}, \citenamefont {Cao},\ and\ \citenamefont {Yan}}]{JinPhysRevLettSkyrmionHall}%
  \BibitemOpen
  \bibfield  {author} {\bibinfo {author} {\bibfnamefont {Z.}~\bibnamefont {Jin}}, \bibinfo {author} {\bibfnamefont {Z.}~\bibnamefont {Zeng}}, \bibinfo {author} {\bibfnamefont {Y.}~\bibnamefont {Cao}},\ and\ \bibinfo {author} {\bibfnamefont {P.}~\bibnamefont {Yan}},\ }\bibfield  {title} {\bibinfo {title} {Skyrmion hall effect in altermagnets},\ }\href {https://doi.org/10.1103/PhysRevLett.133.196701} {\bibfield  {journal} {\bibinfo  {journal} {Phys. Rev. Lett.}\ }\textbf {\bibinfo {volume} {133}},\ \bibinfo {pages} {196701} (\bibinfo {year} {2024})}\BibitemShut {NoStop}%
\bibitem [{\citenamefont {Shao}\ \emph {et~al.}(2021)\citenamefont {Shao}, \citenamefont {Zhang}, \citenamefont {Li}, \citenamefont {Eom},\ and\ \citenamefont {Tsymbal}}]{Shao2021-ge}%
  \BibitemOpen
  \bibfield  {author} {\bibinfo {author} {\bibfnamefont {D.-F.}\ \bibnamefont {Shao}}, \bibinfo {author} {\bibfnamefont {S.-H.}\ \bibnamefont {Zhang}}, \bibinfo {author} {\bibfnamefont {M.}~\bibnamefont {Li}}, \bibinfo {author} {\bibfnamefont {C.-B.}\ \bibnamefont {Eom}},\ and\ \bibinfo {author} {\bibfnamefont {E.~Y.}\ \bibnamefont {Tsymbal}},\ }\bibfield  {title} {\bibinfo {title} {Spin-neutral currents for spintronics},\ }\href {https://doi.org/10.1038/s41467-021-26915-3} {\bibfield  {journal} {\bibinfo  {journal} {Nat. Commun.}\ }\textbf {\bibinfo {volume} {12}},\ \bibinfo {pages} {7061} (\bibinfo {year} {2021})}\BibitemShut {NoStop}%
\bibitem [{\citenamefont {\ifmmode~\check{S}\else \v{S}\fi{}mejkal}\ \emph {et~al.}(2022{\natexlab{c}})\citenamefont {\ifmmode~\check{S}\else \v{S}\fi{}mejkal}, \citenamefont {Hellenes}, \citenamefont {Gonz\'alez-Hern\'andez}, \citenamefont {Sinova},\ and\ \citenamefont {Jungwirth}}]{PhysRevX.12.011028}%
  \BibitemOpen
  \bibfield  {author} {\bibinfo {author} {\bibfnamefont {L.}~\bibnamefont {\ifmmode~\check{S}\else \v{S}\fi{}mejkal}}, \bibinfo {author} {\bibfnamefont {A.~B.}\ \bibnamefont {Hellenes}}, \bibinfo {author} {\bibfnamefont {R.}~\bibnamefont {Gonz\'alez-Hern\'andez}}, \bibinfo {author} {\bibfnamefont {J.}~\bibnamefont {Sinova}},\ and\ \bibinfo {author} {\bibfnamefont {T.}~\bibnamefont {Jungwirth}},\ }\bibfield  {title} {\bibinfo {title} {Giant and tunneling magnetoresistance in unconventional collinear antiferromagnets with nonrelativistic spin-momentum coupling},\ }\href {https://doi.org/10.1103/PhysRevX.12.011028} {\bibfield  {journal} {\bibinfo  {journal} {Phys. Rev. X}\ }\textbf {\bibinfo {volume} {12}},\ \bibinfo {pages} {011028} (\bibinfo {year} {2022}{\natexlab{c}})}\BibitemShut {NoStop}%
\bibitem [{\citenamefont {Reichlova}\ \emph {et~al.}(2024)\citenamefont {Reichlova}, \citenamefont {Lopes~Seeger}, \citenamefont {{Gonz{\'a}lez-Hern{\'a}ndez}}, \citenamefont {Kounta}, \citenamefont {Schlitz}, \citenamefont {Kriegner}, \citenamefont {Ritzinger}, \citenamefont {Lammel}, \citenamefont {Leivisk{\"a}}, \citenamefont {Birk~Hellenes}, \citenamefont {Olejn{\'i}k}, \citenamefont {Pet{\v r}i{\v c}ek}, \citenamefont {Dole{\v z}al}, \citenamefont {Horak}, \citenamefont {Schmoranzerova}, \citenamefont {Badura}, \citenamefont {Bertaina}, \citenamefont {Thomas}, \citenamefont {Baltz}, \citenamefont {Michez}, \citenamefont {Sinova}, \citenamefont {Goennenwein}, \citenamefont {Jungwirth},\ and\ \citenamefont {{\v S}mejkal}}]{reichlovaObservationSpontaneousAnomalous2024a}%
  \BibitemOpen
  \bibfield  {author} {\bibinfo {author} {\bibfnamefont {H.}~\bibnamefont {Reichlova}}, \bibinfo {author} {\bibfnamefont {R.}~\bibnamefont {Lopes~Seeger}}, \bibinfo {author} {\bibfnamefont {R.}~\bibnamefont {{Gonz{\'a}lez-Hern{\'a}ndez}}}, \bibinfo {author} {\bibfnamefont {I.}~\bibnamefont {Kounta}}, \bibinfo {author} {\bibfnamefont {R.}~\bibnamefont {Schlitz}}, \bibinfo {author} {\bibfnamefont {D.}~\bibnamefont {Kriegner}}, \bibinfo {author} {\bibfnamefont {P.}~\bibnamefont {Ritzinger}}, \bibinfo {author} {\bibfnamefont {M.}~\bibnamefont {Lammel}}, \bibinfo {author} {\bibfnamefont {M.}~\bibnamefont {Leivisk{\"a}}}, \bibinfo {author} {\bibfnamefont {A.}~\bibnamefont {Birk~Hellenes}}, \bibinfo {author} {\bibfnamefont {K.}~\bibnamefont {Olejn{\'i}k}}, \bibinfo {author} {\bibfnamefont {V.}~\bibnamefont {Pet{\v r}i{\v c}ek}}, \bibinfo {author} {\bibfnamefont {P.}~\bibnamefont {Dole{\v z}al}}, \bibinfo {author} {\bibfnamefont {L.}~\bibnamefont {Horak}}, \bibinfo {author} {\bibfnamefont {E.}~\bibnamefont
  {Schmoranzerova}}, \bibinfo {author} {\bibfnamefont {A.}~\bibnamefont {Badura}}, \bibinfo {author} {\bibfnamefont {S.}~\bibnamefont {Bertaina}}, \bibinfo {author} {\bibfnamefont {A.}~\bibnamefont {Thomas}}, \bibinfo {author} {\bibfnamefont {V.}~\bibnamefont {Baltz}}, \bibinfo {author} {\bibfnamefont {L.}~\bibnamefont {Michez}}, \bibinfo {author} {\bibfnamefont {J.}~\bibnamefont {Sinova}}, \bibinfo {author} {\bibfnamefont {S.~T.~B.}\ \bibnamefont {Goennenwein}}, \bibinfo {author} {\bibfnamefont {T.}~\bibnamefont {Jungwirth}},\ and\ \bibinfo {author} {\bibfnamefont {L.}~\bibnamefont {{\v S}mejkal}},\ }\bibfield  {title} {\bibinfo {title} {Observation of a spontaneous anomalous hall response in the mn{$_5$}si{$_3$} d-wave altermagnet candidate},\ }\href {https://doi.org/10.1038/s41467-024-48493-w} {\bibfield  {journal} {\bibinfo  {journal} {Nat. Commun.}\ }\textbf {\bibinfo {volume} {15}},\ \bibinfo {pages} {4961} (\bibinfo {year} {2024})}\BibitemShut {NoStop}%
\bibitem [{\citenamefont {Lin}\ \emph {et~al.}(2025)\citenamefont {Lin}, \citenamefont {Zhang}, \citenamefont {Lu},\ and\ \citenamefont {Xie}}]{linCoulombDragAltermagnets2024a}%
  \BibitemOpen
  \bibfield  {author} {\bibinfo {author} {\bibfnamefont {H.-J.}\ \bibnamefont {Lin}}, \bibinfo {author} {\bibfnamefont {S.-B.}\ \bibnamefont {Zhang}}, \bibinfo {author} {\bibfnamefont {H.-Z.}\ \bibnamefont {Lu}},\ and\ \bibinfo {author} {\bibfnamefont {X.~C.}\ \bibnamefont {Xie}},\ }\bibfield  {title} {\bibinfo {title} {Coulomb drag in altermagnets},\ }\href {https://doi.org/10.1103/PhysRevLett.134.136301} {\bibfield  {journal} {\bibinfo  {journal} {Phys. Rev. Lett.}\ }\textbf {\bibinfo {volume} {134}},\ \bibinfo {pages} {136301} (\bibinfo {year} {2025})}\BibitemShut {NoStop}%
\bibitem [{\citenamefont {Ouassou}\ \emph {et~al.}(2023)\citenamefont {Ouassou}, \citenamefont {Brataas},\ and\ \citenamefont {Linder}}]{ouassouDcJosephsonEffect2023a}%
  \BibitemOpen
  \bibfield  {author} {\bibinfo {author} {\bibfnamefont {J.~A.}\ \bibnamefont {Ouassou}}, \bibinfo {author} {\bibfnamefont {A.}~\bibnamefont {Brataas}},\ and\ \bibinfo {author} {\bibfnamefont {J.}~\bibnamefont {Linder}},\ }\bibfield  {title} {\bibinfo {title} {Dc {{Josephson Effect}} in {{Altermagnets}}},\ }\href {https://doi.org/10.1103/PhysRevLett.131.076003} {\bibfield  {journal} {\bibinfo  {journal} {Phys. Rev. Lett.}\ }\textbf {\bibinfo {volume} {131}},\ \bibinfo {pages} {076003} (\bibinfo {year} {2023})}\BibitemShut {NoStop}%
\bibitem [{\citenamefont {Zhu}\ \emph {et~al.}(2025{\natexlab{b}})\citenamefont {Zhu}, \citenamefont {Huang}, \citenamefont {Chen}, \citenamefont {Duan}, \citenamefont {Zhang}, \citenamefont {Zutic},\ and\ \citenamefont {Zhou}}]{zhu2025altermagneticproximityeffect}%
  \BibitemOpen
  \bibfield  {author} {\bibinfo {author} {\bibfnamefont {Z.}~\bibnamefont {Zhu}}, \bibinfo {author} {\bibfnamefont {R.}~\bibnamefont {Huang}}, \bibinfo {author} {\bibfnamefont {X.}~\bibnamefont {Chen}}, \bibinfo {author} {\bibfnamefont {X.}~\bibnamefont {Duan}}, \bibinfo {author} {\bibfnamefont {J.}~\bibnamefont {Zhang}}, \bibinfo {author} {\bibfnamefont {I.}~\bibnamefont {Zutic}},\ and\ \bibinfo {author} {\bibfnamefont {T.}~\bibnamefont {Zhou}},\ }\bibfield  {title} {\bibinfo {title} {Altermagnetic proximity effect},\ }\href {https://arxiv.org/abs/2509.06790} {\bibfield  {journal} {\bibinfo  {journal} {arXiv:2509.06790}\ } (\bibinfo {year} {2025}{\natexlab{b}})}\BibitemShut {NoStop}%
\bibitem [{zhu(2025)}]{zhu2025altermagnetoelectricspinfieldeffect}%
  \BibitemOpen
  \bibfield  {title} {\bibinfo {title} {Altermagnetoelectric spin field effect transistor},\ }\href {https://arxiv.org/abs/2512.02974} {\bibfield  {journal} {\bibinfo  {journal} {arXiv:2512.02974}\ } (\bibinfo {year} {2025})}\BibitemShut {NoStop}%
\bibitem [{\citenamefont {Li}\ and\ \citenamefont {Liu}(2023)}]{liMajoranaCornerModes2023a}%
  \BibitemOpen
  \bibfield  {author} {\bibinfo {author} {\bibfnamefont {Y.-X.}\ \bibnamefont {Li}}\ and\ \bibinfo {author} {\bibfnamefont {C.-C.}\ \bibnamefont {Liu}},\ }\bibfield  {title} {\bibinfo {title} {Majorana corner modes and tunable patterns in an altermagnet heterostructure},\ }\href {https://doi.org/10.1103/PhysRevB.108.205410} {\bibfield  {journal} {\bibinfo  {journal} {Phys. Rev. B}\ }\textbf {\bibinfo {volume} {108}},\ \bibinfo {pages} {205410} (\bibinfo {year} {2023})}\BibitemShut {NoStop}%
\bibitem [{\citenamefont {Ghorashi}\ \emph {et~al.}(2024{\natexlab{a}})\citenamefont {Ghorashi}, \citenamefont {Hughes},\ and\ \citenamefont {Cano}}]{ghorashiAltermagneticRoutesMajorana2024a}%
  \BibitemOpen
  \bibfield  {author} {\bibinfo {author} {\bibfnamefont {S.~A.~A.}\ \bibnamefont {Ghorashi}}, \bibinfo {author} {\bibfnamefont {T.~L.}\ \bibnamefont {Hughes}},\ and\ \bibinfo {author} {\bibfnamefont {J.}~\bibnamefont {Cano}},\ }\bibfield  {title} {\bibinfo {title} {Altermagnetic routes to majorana modes in zero net magnetization},\ }\href {https://doi.org/10.1103/PhysRevLett.133.106601} {\bibfield  {journal} {\bibinfo  {journal} {Phys. Rev. Lett.}\ }\textbf {\bibinfo {volume} {133}},\ \bibinfo {pages} {106601} (\bibinfo {year} {2024}{\natexlab{a}})}\BibitemShut {NoStop}%
\bibitem [{\citenamefont {Li}\ \emph {et~al.}(2024)\citenamefont {Li}, \citenamefont {Liu},\ and\ \citenamefont {Liu}}]{liCreationManipulationHigherorder2024a}%
  \BibitemOpen
  \bibfield  {author} {\bibinfo {author} {\bibfnamefont {Y.-X.}\ \bibnamefont {Li}}, \bibinfo {author} {\bibfnamefont {Y.}~\bibnamefont {Liu}},\ and\ \bibinfo {author} {\bibfnamefont {C.-C.}\ \bibnamefont {Liu}},\ }\bibfield  {title} {\bibinfo {title} {Creation and manipulation of higher-order topological states by altermagnets},\ }\href {https://doi.org/10.1103/PhysRevB.109.L201109} {\bibfield  {journal} {\bibinfo  {journal} {Phys. Rev. B}\ }\textbf {\bibinfo {volume} {109}},\ \bibinfo {pages} {L201109} (\bibinfo {year} {2024})}\BibitemShut {NoStop}%
\bibitem [{\citenamefont {Li}(2024)}]{liRealizingTunableHigherorder2024a}%
  \BibitemOpen
  \bibfield  {author} {\bibinfo {author} {\bibfnamefont {Y.-X.}\ \bibnamefont {Li}},\ }\bibfield  {title} {\bibinfo {title} {Realizing tunable higher-order topological superconductors with altermagnets},\ }\href {https://doi.org/10.1103/PhysRevB.109.224502} {\bibfield  {journal} {\bibinfo  {journal} {Phys. Rev. B}\ }\textbf {\bibinfo {volume} {109}},\ \bibinfo {pages} {224502} (\bibinfo {year} {2024})}\BibitemShut {NoStop}%
\bibitem [{\citenamefont {Li}\ and\ \citenamefont {Zhang}(2025)}]{liFloatingEdgeBands2025}%
  \BibitemOpen
  \bibfield  {author} {\bibinfo {author} {\bibfnamefont {Y.-Y.}\ \bibnamefont {Li}}\ and\ \bibinfo {author} {\bibfnamefont {S.-B.}\ \bibnamefont {Zhang}},\ }\bibfield  {title} {\bibinfo {title} {Floating edge bands in the {{Bernevig-Hughes-Zhang}} model with altermagnetism},\ }\href {https://doi.org/10.1103/PhysRevB.111.045106} {\bibfield  {journal} {\bibinfo  {journal} {Phys. Rev. B}\ }\textbf {\bibinfo {volume} {111}},\ \bibinfo {pages} {045106} (\bibinfo {year} {2025})}\BibitemShut {NoStop}%
\bibitem [{\citenamefont {Rudner}\ and\ \citenamefont {Levitov}(2009)}]{rudner_topological_2009}%
  \BibitemOpen
  \bibfield  {author} {\bibinfo {author} {\bibfnamefont {M.~S.}\ \bibnamefont {Rudner}}\ and\ \bibinfo {author} {\bibfnamefont {L.~S.}\ \bibnamefont {Levitov}},\ }\bibfield  {title} {\bibinfo {title} {Topological transition in a non-{H}ermitian quantum walk},\ }\href {https://doi.org/10.1103/PhysRevLett.102.065703} {\bibfield  {journal} {\bibinfo  {journal} {Phys. Rev. Lett.}\ }\textbf {\bibinfo {volume} {102}},\ \bibinfo {pages} {065703} (\bibinfo {year} {2009})}\BibitemShut {NoStop}%
\bibitem [{\citenamefont {Hu}\ and\ \citenamefont {Hughes}(2011)}]{hu_absence_2011}%
  \BibitemOpen
  \bibfield  {author} {\bibinfo {author} {\bibfnamefont {Y.~C.}\ \bibnamefont {Hu}}\ and\ \bibinfo {author} {\bibfnamefont {T.~L.}\ \bibnamefont {Hughes}},\ }\bibfield  {title} {\bibinfo {title} {Absence of topological insulator phases in non-{H}ermitian {PT}-symmetric {H}amiltonians},\ }\href {https://doi.org/10.1103/PhysRevB.84.153101} {\bibfield  {journal} {\bibinfo  {journal} {Phys. Rev. B}\ }\textbf {\bibinfo {volume} {84}},\ \bibinfo {pages} {153101} (\bibinfo {year} {2011})}\BibitemShut {NoStop}%
\bibitem [{\citenamefont {Lee}(2016)}]{lee_anomalous_2016}%
  \BibitemOpen
  \bibfield  {author} {\bibinfo {author} {\bibfnamefont {T.~E.}\ \bibnamefont {Lee}},\ }\bibfield  {title} {\bibinfo {title} {Anomalous edge state in a non-{H}ermitian lattice},\ }\href {https://doi.org/10.1103/PhysRevLett.116.133903} {\bibfield  {journal} {\bibinfo  {journal} {Phys. Rev. Lett.}\ }\textbf {\bibinfo {volume} {116}},\ \bibinfo {pages} {133903} (\bibinfo {year} {2016})}\BibitemShut {NoStop}%
\bibitem [{\citenamefont {Shen}\ \emph {et~al.}(2018)\citenamefont {Shen}, \citenamefont {Zhen},\ and\ \citenamefont {Fu}}]{shen_topological_2018}%
  \BibitemOpen
  \bibfield  {author} {\bibinfo {author} {\bibfnamefont {H.}~\bibnamefont {Shen}}, \bibinfo {author} {\bibfnamefont {B.}~\bibnamefont {Zhen}},\ and\ \bibinfo {author} {\bibfnamefont {L.}~\bibnamefont {Fu}},\ }\bibfield  {title} {\bibinfo {title} {Topological band theory for non-{H}ermitian {H}amiltonians},\ }\href {https://doi.org/10.1103/PhysRevLett.120.146402} {\bibfield  {journal} {\bibinfo  {journal} {Phys. Rev. Lett.}\ }\textbf {\bibinfo {volume} {120}},\ \bibinfo {pages} {146402} (\bibinfo {year} {2018})}\BibitemShut {NoStop}%
\bibitem [{\citenamefont {Yao}\ and\ \citenamefont {Wang}(2018)}]{yao_edge_2018}%
  \BibitemOpen
  \bibfield  {author} {\bibinfo {author} {\bibfnamefont {S.}~\bibnamefont {Yao}}\ and\ \bibinfo {author} {\bibfnamefont {Z.}~\bibnamefont {Wang}},\ }\bibfield  {title} {\bibinfo {title} {Edge states and topological invariants of non-{H}ermitian systems},\ }\href {https://doi.org/10.1103/PhysRevLett.121.086803} {\bibfield  {journal} {\bibinfo  {journal} {Phys. Rev. Lett.}\ }\textbf {\bibinfo {volume} {121}},\ \bibinfo {pages} {086803} (\bibinfo {year} {2018})}\BibitemShut {NoStop}%
\bibitem [{\citenamefont {Yao}\ \emph {et~al.}(2018)\citenamefont {Yao}, \citenamefont {Song},\ and\ \citenamefont {Wang}}]{yao_nonhermitian_2018}%
  \BibitemOpen
  \bibfield  {author} {\bibinfo {author} {\bibfnamefont {S.}~\bibnamefont {Yao}}, \bibinfo {author} {\bibfnamefont {F.}~\bibnamefont {Song}},\ and\ \bibinfo {author} {\bibfnamefont {Z.}~\bibnamefont {Wang}},\ }\bibfield  {title} {\bibinfo {title} {Non-{H}ermitian {C}hern bands},\ }\href {https://doi.org/10.1103/PhysRevLett.121.136802} {\bibfield  {journal} {\bibinfo  {journal} {Phys. Rev. Lett.}\ }\textbf {\bibinfo {volume} {121}},\ \bibinfo {pages} {136802} (\bibinfo {year} {2018})}\BibitemShut {NoStop}%
\bibitem [{\citenamefont {Kunst}\ \emph {et~al.}(2018)\citenamefont {Kunst}, \citenamefont {Edvardsson}, \citenamefont {Budich},\ and\ \citenamefont {Bergholtz}}]{kunst_biorthogonal_2018}%
  \BibitemOpen
  \bibfield  {author} {\bibinfo {author} {\bibfnamefont {F.~K.}\ \bibnamefont {Kunst}}, \bibinfo {author} {\bibfnamefont {E.}~\bibnamefont {Edvardsson}}, \bibinfo {author} {\bibfnamefont {J.~C.}\ \bibnamefont {Budich}},\ and\ \bibinfo {author} {\bibfnamefont {E.~J.}\ \bibnamefont {Bergholtz}},\ }\bibfield  {title} {\bibinfo {title} {Biorthogonal bulk-boundary correspondence in non-{H}ermitian systems},\ }\href {https://doi.org/10.1103/PhysRevLett.121.026808} {\bibfield  {journal} {\bibinfo  {journal} {Phys. Rev. Lett.}\ }\textbf {\bibinfo {volume} {121}},\ \bibinfo {pages} {026808} (\bibinfo {year} {2018})}\BibitemShut {NoStop}%
\bibitem [{\citenamefont {Koutserimpas}\ and\ \citenamefont {Fleury}(2018)}]{koutserimpas_nonreciprocal_2018}%
  \BibitemOpen
  \bibfield  {author} {\bibinfo {author} {\bibfnamefont {T.~T.}\ \bibnamefont {Koutserimpas}}\ and\ \bibinfo {author} {\bibfnamefont {R.}~\bibnamefont {Fleury}},\ }\bibfield  {title} {\bibinfo {title} {Nonreciprocal gain in non-{H}ermitian time-{F}loquet systems},\ }\href {https://doi.org/10.1103/PhysRevLett.120.087401} {\bibfield  {journal} {\bibinfo  {journal} {Phys. Rev. Lett.}\ }\textbf {\bibinfo {volume} {120}},\ \bibinfo {pages} {087401} (\bibinfo {year} {2018})}\BibitemShut {NoStop}%
\bibitem [{\citenamefont {Longhi}(2019)}]{longhi_topological_2019}%
  \BibitemOpen
  \bibfield  {author} {\bibinfo {author} {\bibfnamefont {S.}~\bibnamefont {Longhi}},\ }\bibfield  {title} {\bibinfo {title} {Topological phase transition in non-{H}ermitian quasicrystals},\ }\href {https://doi.org/10.1103/PhysRevLett.122.237601} {\bibfield  {journal} {\bibinfo  {journal} {Phys. Rev. Lett.}\ }\textbf {\bibinfo {volume} {122}},\ \bibinfo {pages} {237601} (\bibinfo {year} {2019})}\BibitemShut {NoStop}%
\bibitem [{\citenamefont {Liu}\ \emph {et~al.}(2019)\citenamefont {Liu}, \citenamefont {Zhang}, \citenamefont {Ai}, \citenamefont {Gong}, \citenamefont {Kawabata}, \citenamefont {Ueda},\ and\ \citenamefont {Nori}}]{liu_secondorder_2019}%
  \BibitemOpen
  \bibfield  {author} {\bibinfo {author} {\bibfnamefont {T.}~\bibnamefont {Liu}}, \bibinfo {author} {\bibfnamefont {Y.-R.}\ \bibnamefont {Zhang}}, \bibinfo {author} {\bibfnamefont {Q.}~\bibnamefont {Ai}}, \bibinfo {author} {\bibfnamefont {Z.}~\bibnamefont {Gong}}, \bibinfo {author} {\bibfnamefont {K.}~\bibnamefont {Kawabata}}, \bibinfo {author} {\bibfnamefont {M.}~\bibnamefont {Ueda}},\ and\ \bibinfo {author} {\bibfnamefont {F.}~\bibnamefont {Nori}},\ }\bibfield  {title} {\bibinfo {title} {Second-order topological phases in non-{H}ermitian systems},\ }\href {https://doi.org/10.1103/PhysRevLett.122.076801} {\bibfield  {journal} {\bibinfo  {journal} {Phys. Rev. Lett.}\ }\textbf {\bibinfo {volume} {122}},\ \bibinfo {pages} {076801} (\bibinfo {year} {2019})}\BibitemShut {NoStop}%
\bibitem [{\citenamefont {Luo}\ and\ \citenamefont {Zhang}(2019)}]{luo_higherorder_2019}%
  \BibitemOpen
  \bibfield  {author} {\bibinfo {author} {\bibfnamefont {X.-W.}\ \bibnamefont {Luo}}\ and\ \bibinfo {author} {\bibfnamefont {C.}~\bibnamefont {Zhang}},\ }\bibfield  {title} {\bibinfo {title} {Higher-order topological corner states induced by gain and loss},\ }\href {https://doi.org/10.1103/PhysRevLett.123.073601} {\bibfield  {journal} {\bibinfo  {journal} {Phys. Rev. Lett.}\ }\textbf {\bibinfo {volume} {123}},\ \bibinfo {pages} {073601} (\bibinfo {year} {2019})}\BibitemShut {NoStop}%
\bibitem [{\citenamefont {Matsumoto}\ \emph {et~al.}(2020)\citenamefont {Matsumoto}, \citenamefont {Kawabata}, \citenamefont {Ashida}, \citenamefont {Furukawa},\ and\ \citenamefont {Ueda}}]{matsumoto_continuous_2020}%
  \BibitemOpen
  \bibfield  {author} {\bibinfo {author} {\bibfnamefont {N.}~\bibnamefont {Matsumoto}}, \bibinfo {author} {\bibfnamefont {K.}~\bibnamefont {Kawabata}}, \bibinfo {author} {\bibfnamefont {Y.}~\bibnamefont {Ashida}}, \bibinfo {author} {\bibfnamefont {S.}~\bibnamefont {Furukawa}},\ and\ \bibinfo {author} {\bibfnamefont {M.}~\bibnamefont {Ueda}},\ }\bibfield  {title} {\bibinfo {title} {Continuous phase transition without gap closing in non-{H}ermitian quantum many-body systems},\ }\href {https://doi.org/10.1103/PhysRevLett.125.260601} {\bibfield  {journal} {\bibinfo  {journal} {Phys. Rev. Lett.}\ }\textbf {\bibinfo {volume} {125}},\ \bibinfo {pages} {260601} (\bibinfo {year} {2020})}\BibitemShut {NoStop}%
\bibitem [{\citenamefont {Xiao}\ \emph {et~al.}(2020)\citenamefont {Xiao}, \citenamefont {Deng}, \citenamefont {Wang}, \citenamefont {Zhu}, \citenamefont {Wang}, \citenamefont {Yi},\ and\ \citenamefont {Xue}}]{xiao_nonhermitian_2020}%
  \BibitemOpen
  \bibfield  {author} {\bibinfo {author} {\bibfnamefont {L.}~\bibnamefont {Xiao}}, \bibinfo {author} {\bibfnamefont {T.}~\bibnamefont {Deng}}, \bibinfo {author} {\bibfnamefont {K.}~\bibnamefont {Wang}}, \bibinfo {author} {\bibfnamefont {G.}~\bibnamefont {Zhu}}, \bibinfo {author} {\bibfnamefont {Z.}~\bibnamefont {Wang}}, \bibinfo {author} {\bibfnamefont {W.}~\bibnamefont {Yi}},\ and\ \bibinfo {author} {\bibfnamefont {P.}~\bibnamefont {Xue}},\ }\bibfield  {title} {\bibinfo {title} {Non-{H}ermitian bulk--boundary correspondence in quantum dynamics},\ }\href {https://doi.org/10.1038/s41567-020-0836-6} {\bibfield  {journal} {\bibinfo  {journal} {Nat. Phys.}\ }\textbf {\bibinfo {volume} {16}},\ \bibinfo {pages} {761} (\bibinfo {year} {2020})}\BibitemShut {NoStop}%
\bibitem [{\citenamefont {Bergholtz}\ \emph {et~al.}(2021)\citenamefont {Bergholtz}, \citenamefont {Budich},\ and\ \citenamefont {Kunst}}]{bergholtz_exceptional_2021}%
  \BibitemOpen
  \bibfield  {author} {\bibinfo {author} {\bibfnamefont {E.~J.}\ \bibnamefont {Bergholtz}}, \bibinfo {author} {\bibfnamefont {J.~C.}\ \bibnamefont {Budich}},\ and\ \bibinfo {author} {\bibfnamefont {F.~K.}\ \bibnamefont {Kunst}},\ }\bibfield  {title} {\bibinfo {title} {Exceptional topology of non-{H}ermitian systems},\ }\href {https://doi.org/10.1103/RevModPhys.93.015005} {\bibfield  {journal} {\bibinfo  {journal} {Rev. Mod. Phys.}\ }\textbf {\bibinfo {volume} {93}},\ \bibinfo {pages} {015005} (\bibinfo {year} {2021})}\BibitemShut {NoStop}%
\bibitem [{\citenamefont {Ding}\ \emph {et~al.}(2022)\citenamefont {Ding}, \citenamefont {Fang},\ and\ \citenamefont {Ma}}]{ding_nonhermitian_2022}%
  \BibitemOpen
  \bibfield  {author} {\bibinfo {author} {\bibfnamefont {K.}~\bibnamefont {Ding}}, \bibinfo {author} {\bibfnamefont {C.}~\bibnamefont {Fang}},\ and\ \bibinfo {author} {\bibfnamefont {G.}~\bibnamefont {Ma}},\ }\bibfield  {title} {\bibinfo {title} {Non-{H}ermitian topology and exceptional-point geometries},\ }\href {https://doi.org/10.1038/s42254-022-00516-5} {\bibfield  {journal} {\bibinfo  {journal} {Nat. Rev. Phys.}\ }\textbf {\bibinfo {volume} {4}},\ \bibinfo {pages} {745} (\bibinfo {year} {2022})}\BibitemShut {NoStop}%
\bibitem [{\citenamefont {Li}\ \emph {et~al.}(2022{\natexlab{a}})\citenamefont {Li}, \citenamefont {Ji}, \citenamefont {Chen}, \citenamefont {Yan},\ and\ \citenamefont {Yang}}]{li_topological_2022}%
  \BibitemOpen
  \bibfield  {author} {\bibinfo {author} {\bibfnamefont {Y.}~\bibnamefont {Li}}, \bibinfo {author} {\bibfnamefont {X.}~\bibnamefont {Ji}}, \bibinfo {author} {\bibfnamefont {Y.}~\bibnamefont {Chen}}, \bibinfo {author} {\bibfnamefont {X.}~\bibnamefont {Yan}},\ and\ \bibinfo {author} {\bibfnamefont {X.}~\bibnamefont {Yang}},\ }\bibfield  {title} {\bibinfo {title} {Topological energy braiding of non-{B}loch bands},\ }\href {https://doi.org/10.1103/PhysRevB.106.195425} {\bibfield  {journal} {\bibinfo  {journal} {Phys. Rev. B}\ }\textbf {\bibinfo {volume} {106}},\ \bibinfo {pages} {195425} (\bibinfo {year} {2022}{\natexlab{a}})}\BibitemShut {NoStop}%
\bibitem [{\citenamefont {Zhang}\ \emph {et~al.}(2022{\natexlab{a}})\citenamefont {Zhang}, \citenamefont {Zhang}, \citenamefont {Lu},\ and\ \citenamefont {Chen}}]{zhang_review_2022}%
  \BibitemOpen
  \bibfield  {author} {\bibinfo {author} {\bibfnamefont {X.}~\bibnamefont {Zhang}}, \bibinfo {author} {\bibfnamefont {T.}~\bibnamefont {Zhang}}, \bibinfo {author} {\bibfnamefont {M.-H.}\ \bibnamefont {Lu}},\ and\ \bibinfo {author} {\bibfnamefont {Y.-F.}\ \bibnamefont {Chen}},\ }\bibfield  {title} {\bibinfo {title} {A review on non-{H}ermitian skin effect},\ }\href {https://doi.org/10.1080/23746149.2022.2109431} {\bibfield  {journal} {\bibinfo  {journal} {Advances in Physics: X}\ }\textbf {\bibinfo {volume} {7}},\ \bibinfo {pages} {2109431} (\bibinfo {year} {2022}{\natexlab{a}})}\BibitemShut {NoStop}%
\bibitem [{\citenamefont {Weidemann}\ \emph {et~al.}(2022)\citenamefont {Weidemann}, \citenamefont {Kremer}, \citenamefont {Longhi},\ and\ \citenamefont {Szameit}}]{weidemann_topological_2022}%
  \BibitemOpen
  \bibfield  {author} {\bibinfo {author} {\bibfnamefont {S.}~\bibnamefont {Weidemann}}, \bibinfo {author} {\bibfnamefont {M.}~\bibnamefont {Kremer}}, \bibinfo {author} {\bibfnamefont {S.}~\bibnamefont {Longhi}},\ and\ \bibinfo {author} {\bibfnamefont {A.}~\bibnamefont {Szameit}},\ }\bibfield  {title} {\bibinfo {title} {Topological triple phase transition in non-{H}ermitian {F}loquet quasicrystals},\ }\href {https://doi.org/10.1038/s41586-021-04253-0} {\bibfield  {journal} {\bibinfo  {journal} {Nature}\ }\textbf {\bibinfo {volume} {601}},\ \bibinfo {pages} {354} (\bibinfo {year} {2022})}\BibitemShut {NoStop}%
\bibitem [{\citenamefont {Okuma}\ and\ \citenamefont {Sato}(2023)}]{okuma_nonhermitian_2023}%
  \BibitemOpen
  \bibfield  {author} {\bibinfo {author} {\bibfnamefont {N.}~\bibnamefont {Okuma}}\ and\ \bibinfo {author} {\bibfnamefont {M.}~\bibnamefont {Sato}},\ }\bibfield  {title} {\bibinfo {title} {Non-{H}ermitian topological phenomena: A review},\ }\href {https://doi.org/10.1146/annurev-conmatphys-040521-033133} {\bibfield  {journal} {\bibinfo  {journal} {Annu. Rev. Condens. Matter Phys.}\ }\textbf {\bibinfo {volume} {14}},\ \bibinfo {pages} {83} (\bibinfo {year} {2023})}\BibitemShut {NoStop}%
\bibitem [{\citenamefont {Ji}\ and\ \citenamefont {Yang}(2024)}]{ji_generalized_2024}%
  \BibitemOpen
  \bibfield  {author} {\bibinfo {author} {\bibfnamefont {X.}~\bibnamefont {Ji}}\ and\ \bibinfo {author} {\bibfnamefont {X.}~\bibnamefont {Yang}},\ }\bibfield  {title} {\bibinfo {title} {Generalized bulk-boundary correspondence in periodically driven non-{H}ermitian systems},\ }\href {https://doi.org/10.1088/1361-648X/ad2c73} {\bibfield  {journal} {\bibinfo  {journal} {J. Phys.: Condens. Matter}\ }\textbf {\bibinfo {volume} {36}},\ \bibinfo {pages} {243001} (\bibinfo {year} {2024})}\BibitemShut {NoStop}%
\bibitem [{\citenamefont {Wang}\ \emph {et~al.}(2025{\natexlab{b}})\citenamefont {Wang}, \citenamefont {Cheng}, \citenamefont {Zou}, \citenamefont {Ge}, \citenamefont {Zhao}, \citenamefont {Si}, \citenamefont {Yuan}, \citenamefont {Sun}, \citenamefont {Xue},\ and\ \citenamefont {Zhang}}]{pnas2025_wbb}%
  \BibitemOpen
  \bibfield  {author} {\bibinfo {author} {\bibfnamefont {B.-B.}\ \bibnamefont {Wang}}, \bibinfo {author} {\bibfnamefont {Z.}~\bibnamefont {Cheng}}, \bibinfo {author} {\bibfnamefont {H.-Y.}\ \bibnamefont {Zou}}, \bibinfo {author} {\bibfnamefont {Y.}~\bibnamefont {Ge}}, \bibinfo {author} {\bibfnamefont {K.-Q.}\ \bibnamefont {Zhao}}, \bibinfo {author} {\bibfnamefont {Q.-R.}\ \bibnamefont {Si}}, \bibinfo {author} {\bibfnamefont {S.-Q.}\ \bibnamefont {Yuan}}, \bibinfo {author} {\bibfnamefont {H.-X.}\ \bibnamefont {Sun}}, \bibinfo {author} {\bibfnamefont {H.}~\bibnamefont {Xue}},\ and\ \bibinfo {author} {\bibfnamefont {B.}~\bibnamefont {Zhang}},\ }\bibfield  {title} {\bibinfo {title} {Observationf of disorder-induced boundary localization},\ }\href {https://doi.org/10.1073/pnas.2422154122} {\bibfield  {journal} {\bibinfo  {journal} {Proc. Natl. Acad. Sci. U.S.A.}\ }\textbf {\bibinfo {volume} {122}},\ \bibinfo {pages} {e2422154122} (\bibinfo {year} {2025}{\natexlab{b}})}\BibitemShut {NoStop}%
\bibitem [{\citenamefont {Ji}\ \emph {et~al.}(2025)\citenamefont {Ji}, \citenamefont {Geng}, \citenamefont {Akhtar},\ and\ \citenamefont {Yang}}]{ji2025floqueta}%
  \BibitemOpen
  \bibfield  {author} {\bibinfo {author} {\bibfnamefont {X.}~\bibnamefont {Ji}}, \bibinfo {author} {\bibfnamefont {H.}~\bibnamefont {Geng}}, \bibinfo {author} {\bibfnamefont {N.}~\bibnamefont {Akhtar}},\ and\ \bibinfo {author} {\bibfnamefont {X.}~\bibnamefont {Yang}},\ }\bibfield  {title} {\bibinfo {title} {Floquet engineering of point-gapped topological superconductors},\ }\href {https://doi.org/10.1103/PhysRevB.111.195419} {\bibfield  {journal} {\bibinfo  {journal} {Phys. Rev. B}\ }\textbf {\bibinfo {volume} {111}},\ \bibinfo {pages} {195419} (\bibinfo {year} {2025})}\BibitemShut {NoStop}%
\bibitem [{\citenamefont {Bender}\ and\ \citenamefont {Boettcher}(1998)}]{bender_real_1998}%
  \BibitemOpen
  \bibfield  {author} {\bibinfo {author} {\bibfnamefont {C.~M.}\ \bibnamefont {Bender}}\ and\ \bibinfo {author} {\bibfnamefont {S.}~\bibnamefont {Boettcher}},\ }\bibfield  {title} {\bibinfo {title} {Real spectra in non-{H}ermitian {H}amiltonians having $\mathcal{PT}$ symmetry},\ }\href {https://doi.org/10.1103/PhysRevLett.80.5243} {\bibfield  {journal} {\bibinfo  {journal} {Phys. Rev. Lett.}\ }\textbf {\bibinfo {volume} {80}},\ \bibinfo {pages} {5243} (\bibinfo {year} {1998})}\BibitemShut {NoStop}%
\bibitem [{\citenamefont {Hodaei}\ \emph {et~al.}(2017)\citenamefont {Hodaei}, \citenamefont {Hassan}, \citenamefont {Wittek}, \citenamefont {{Garcia-Gracia}}, \citenamefont {{El-Ganainy}}, \citenamefont {Christodoulides},\ and\ \citenamefont {Khajavikhan}}]{hodaei_enhanced_2017}%
  \BibitemOpen
  \bibfield  {author} {\bibinfo {author} {\bibfnamefont {H.}~\bibnamefont {Hodaei}}, \bibinfo {author} {\bibfnamefont {A.~U.}\ \bibnamefont {Hassan}}, \bibinfo {author} {\bibfnamefont {S.}~\bibnamefont {Wittek}}, \bibinfo {author} {\bibfnamefont {H.}~\bibnamefont {{Garcia-Gracia}}}, \bibinfo {author} {\bibfnamefont {R.}~\bibnamefont {{El-Ganainy}}}, \bibinfo {author} {\bibfnamefont {D.~N.}\ \bibnamefont {Christodoulides}},\ and\ \bibinfo {author} {\bibfnamefont {M.}~\bibnamefont {Khajavikhan}},\ }\bibfield  {title} {\bibinfo {title} {Enhanced sensitivity at higher-order exceptional points},\ }\href {https://doi.org/10.1038/nature23280} {\bibfield  {journal} {\bibinfo  {journal} {Nature}\ }\textbf {\bibinfo {volume} {548}},\ \bibinfo {pages} {187} (\bibinfo {year} {2017})}\BibitemShut {NoStop}%
\bibitem [{\citenamefont {Ashida}\ \emph {et~al.}(2020)\citenamefont {Ashida}, \citenamefont {Gong},\ and\ \citenamefont {Ueda}}]{ashida_nonhermitian_2020}%
  \BibitemOpen
  \bibfield  {author} {\bibinfo {author} {\bibfnamefont {Y.}~\bibnamefont {Ashida}}, \bibinfo {author} {\bibfnamefont {Z.}~\bibnamefont {Gong}},\ and\ \bibinfo {author} {\bibfnamefont {M.}~\bibnamefont {Ueda}},\ }\bibfield  {title} {\bibinfo {title} {Non-{H}ermitian physics},\ }\href {https://doi.org/10.1080/00018732.2021.1876991} {\bibfield  {journal} {\bibinfo  {journal} {Advances in Physics}\ }\textbf {\bibinfo {volume} {69}},\ \bibinfo {pages} {249} (\bibinfo {year} {2020})}\BibitemShut {NoStop}%
\bibitem [{\citenamefont {Lin}\ \emph {et~al.}(2023)\citenamefont {Lin}, \citenamefont {Tai}, \citenamefont {Yang}, \citenamefont {Li},\ and\ \citenamefont {Lee}}]{lin_topological_2023}%
  \BibitemOpen
  \bibfield  {author} {\bibinfo {author} {\bibfnamefont {R.}~\bibnamefont {Lin}}, \bibinfo {author} {\bibfnamefont {T.}~\bibnamefont {Tai}}, \bibinfo {author} {\bibfnamefont {M.}~\bibnamefont {Yang}}, \bibinfo {author} {\bibfnamefont {L.}~\bibnamefont {Li}},\ and\ \bibinfo {author} {\bibfnamefont {C.~H.}\ \bibnamefont {Lee}},\ }\bibfield  {title} {\bibinfo {title} {Topological non-{H}ermitian skin effect},\ }\href {https://doi.org/10.1007/s11467-023-1309-z} {\bibfield  {journal} {\bibinfo  {journal} {Front. Phys.}\ }\textbf {\bibinfo {volume} {18}},\ \bibinfo {pages} {53605} (\bibinfo {year} {2023})}\BibitemShut {NoStop}%
\bibitem [{\citenamefont {Wang}\ \emph {et~al.}(2024)\citenamefont {Wang}, \citenamefont {Zhong},\ and\ \citenamefont {Fan}}]{wang_nonhermitian_2024}%
  \BibitemOpen
  \bibfield  {author} {\bibinfo {author} {\bibfnamefont {H.}~\bibnamefont {Wang}}, \bibinfo {author} {\bibfnamefont {J.}~\bibnamefont {Zhong}},\ and\ \bibinfo {author} {\bibfnamefont {S.}~\bibnamefont {Fan}},\ }\bibfield  {title} {\bibinfo {title} {Non-{H}ermitian photonic band winding and skin effects: A tutorial},\ }\href {https://doi.org/10.1364/AOP.529289} {\bibfield  {journal} {\bibinfo  {journal} {Adv. Opt. Photon.}\ }\textbf {\bibinfo {volume} {16}},\ \bibinfo {pages} {659} (\bibinfo {year} {2024})}\BibitemShut {NoStop}%
\bibitem [{\citenamefont {Miri}\ and\ \citenamefont {Al{\textasciidieresis}{\textasciiacute}}(2019)}]{miri_exceptional_2019}%
  \BibitemOpen
  \bibfield  {author} {\bibinfo {author} {\bibfnamefont {M.-A.}\ \bibnamefont {Miri}}\ and\ \bibinfo {author} {\bibfnamefont {A.}~\bibnamefont {Al{\textasciidieresis}{\textasciiacute}}},\ }\bibfield  {title} {\bibinfo {title} {Exceptional points in optics and photonics},\ }\href {https://doi.org/10.1126/science.aar7709} {\bibfield  {journal} {\bibinfo  {journal} {Science}\ }\textbf {\bibinfo {volume} {363}},\ \bibinfo {pages} {eaar7709} (\bibinfo {year} {2019})}\BibitemShut {NoStop}%
\bibitem [{\citenamefont {Wang}\ and\ \citenamefont {Yan}(2025)}]{shuxuanprbep}%
  \BibitemOpen
  \bibfield  {author} {\bibinfo {author} {\bibfnamefont {S.-X.}\ \bibnamefont {Wang}}\ and\ \bibinfo {author} {\bibfnamefont {Z.}~\bibnamefont {Yan}},\ }\bibfield  {title} {\bibinfo {title} {Theory for the spectral splitting exponent of exceptional points},\ }\href {https://doi.org/10.1103/nrg4-gl5z} {\bibfield  {journal} {\bibinfo  {journal} {Phys. Rev. B}\ }\textbf {\bibinfo {volume} {112}},\ \bibinfo {pages} {195125} (\bibinfo {year} {2025})}\BibitemShut {NoStop}%
\bibitem [{\citenamefont {Alvarez}\ \emph {et~al.}(2018)\citenamefont {Alvarez}, \citenamefont {Vargas},\ and\ \citenamefont {Torres}}]{alvarez_nonhermitian_2018}%
  \BibitemOpen
  \bibfield  {author} {\bibinfo {author} {\bibfnamefont {V.~M.}\ \bibnamefont {Alvarez}}, \bibinfo {author} {\bibfnamefont {J.~B.}\ \bibnamefont {Vargas}},\ and\ \bibinfo {author} {\bibfnamefont {L.~F.}\ \bibnamefont {Torres}},\ }\bibfield  {title} {\bibinfo {title} {Non-{H}ermitian robust edge states in one dimension: Anomalous localization and eigenspace condensation at exceptional points},\ }\href {https://doi.org/10.1103/PhysRevB.97.121401} {\bibfield  {journal} {\bibinfo  {journal} {Phys. Rev. B}\ }\textbf {\bibinfo {volume} {97}},\ \bibinfo {pages} {121401} (\bibinfo {year} {2018})}\BibitemShut {NoStop}%
\bibitem [{\citenamefont {Song}\ \emph {et~al.}(2019{\natexlab{a}})\citenamefont {Song}, \citenamefont {Yao},\ and\ \citenamefont {Wang}}]{song_nonhermitian_2019a}%
  \BibitemOpen
  \bibfield  {author} {\bibinfo {author} {\bibfnamefont {F.}~\bibnamefont {Song}}, \bibinfo {author} {\bibfnamefont {S.}~\bibnamefont {Yao}},\ and\ \bibinfo {author} {\bibfnamefont {Z.}~\bibnamefont {Wang}},\ }\bibfield  {title} {\bibinfo {title} {Non-{H}ermitian topological invariants in real space},\ }\href {https://doi.org/10.1103/PhysRevLett.123.246801} {\bibfield  {journal} {\bibinfo  {journal} {Phys. Rev. Lett.}\ }\textbf {\bibinfo {volume} {123}},\ \bibinfo {pages} {246801} (\bibinfo {year} {2019}{\natexlab{a}})}\BibitemShut {NoStop}%
\bibitem [{\citenamefont {Song}\ \emph {et~al.}(2019{\natexlab{b}})\citenamefont {Song}, \citenamefont {Yao},\ and\ \citenamefont {Wang}}]{song_nonhermitian_2019}%
  \BibitemOpen
  \bibfield  {author} {\bibinfo {author} {\bibfnamefont {F.}~\bibnamefont {Song}}, \bibinfo {author} {\bibfnamefont {S.}~\bibnamefont {Yao}},\ and\ \bibinfo {author} {\bibfnamefont {Z.}~\bibnamefont {Wang}},\ }\bibfield  {title} {\bibinfo {title} {Non-{H}ermitian skin effect and chiral damping in open quantum systems},\ }\href {https://doi.org/10.1103/PhysRevLett.123.170401} {\bibfield  {journal} {\bibinfo  {journal} {Phys. Rev. Lett.}\ }\textbf {\bibinfo {volume} {123}},\ \bibinfo {pages} {170401} (\bibinfo {year} {2019}{\natexlab{b}})}\BibitemShut {NoStop}%
\bibitem [{\citenamefont {Zhang}\ \emph {et~al.}(2020)\citenamefont {Zhang}, \citenamefont {Yang},\ and\ \citenamefont {Fang}}]{zhang_correspondence_2020}%
  \BibitemOpen
  \bibfield  {author} {\bibinfo {author} {\bibfnamefont {K.}~\bibnamefont {Zhang}}, \bibinfo {author} {\bibfnamefont {Z.}~\bibnamefont {Yang}},\ and\ \bibinfo {author} {\bibfnamefont {C.}~\bibnamefont {Fang}},\ }\bibfield  {title} {\bibinfo {title} {Correspondence between winding numbers and skin modes in non-{H}ermitian systems},\ }\href {https://doi.org/10.1103/PhysRevLett.125.126402} {\bibfield  {journal} {\bibinfo  {journal} {Phys. Rev. Lett.}\ }\textbf {\bibinfo {volume} {125}},\ \bibinfo {pages} {126402} (\bibinfo {year} {2020})}\BibitemShut {NoStop}%
\bibitem [{\citenamefont {Okuma}\ \emph {et~al.}(2020)\citenamefont {Okuma}, \citenamefont {Kawabata}, \citenamefont {Shiozaki},\ and\ \citenamefont {Sato}}]{okuma_topological_2020}%
  \BibitemOpen
  \bibfield  {author} {\bibinfo {author} {\bibfnamefont {N.}~\bibnamefont {Okuma}}, \bibinfo {author} {\bibfnamefont {K.}~\bibnamefont {Kawabata}}, \bibinfo {author} {\bibfnamefont {K.}~\bibnamefont {Shiozaki}},\ and\ \bibinfo {author} {\bibfnamefont {M.}~\bibnamefont {Sato}},\ }\bibfield  {title} {\bibinfo {title} {Topological origin of non-{H}ermitian skin effects},\ }\href {https://doi.org/10.1103/PhysRevLett.124.086801} {\bibfield  {journal} {\bibinfo  {journal} {Phys. Rev. Lett.}\ }\textbf {\bibinfo {volume} {124}},\ \bibinfo {pages} {086801} (\bibinfo {year} {2020})}\BibitemShut {NoStop}%
\bibitem [{\citenamefont {Li}\ \emph {et~al.}(2020{\natexlab{a}})\citenamefont {Li}, \citenamefont {Lee}, \citenamefont {Mu},\ and\ \citenamefont {Gong}}]{li_critical_2020}%
  \BibitemOpen
  \bibfield  {author} {\bibinfo {author} {\bibfnamefont {L.}~\bibnamefont {Li}}, \bibinfo {author} {\bibfnamefont {C.~H.}\ \bibnamefont {Lee}}, \bibinfo {author} {\bibfnamefont {S.}~\bibnamefont {Mu}},\ and\ \bibinfo {author} {\bibfnamefont {J.}~\bibnamefont {Gong}},\ }\bibfield  {title} {\bibinfo {title} {Critical non-{H}ermitian skin effect},\ }\href {https://doi.org/10.1038/s41467-020-18917-4} {\bibfield  {journal} {\bibinfo  {journal} {Nat. Commun.}\ }\textbf {\bibinfo {volume} {11}},\ \bibinfo {pages} {5491} (\bibinfo {year} {2020}{\natexlab{a}})}\BibitemShut {NoStop}%
\bibitem [{\citenamefont {Zhang}\ \emph {et~al.}(2022{\natexlab{b}})\citenamefont {Zhang}, \citenamefont {Yang},\ and\ \citenamefont {Fang}}]{zhang_universal_2022}%
  \BibitemOpen
  \bibfield  {author} {\bibinfo {author} {\bibfnamefont {K.}~\bibnamefont {Zhang}}, \bibinfo {author} {\bibfnamefont {Z.}~\bibnamefont {Yang}},\ and\ \bibinfo {author} {\bibfnamefont {C.}~\bibnamefont {Fang}},\ }\bibfield  {title} {\bibinfo {title} {Universal non-{H}ermitian skin effect in two and higher dimensions},\ }\href {https://doi.org/10.1038/s41467-022-30161-6} {\bibfield  {journal} {\bibinfo  {journal} {Nat. Commun.}\ }\textbf {\bibinfo {volume} {13}},\ \bibinfo {pages} {2496} (\bibinfo {year} {2022}{\natexlab{b}})}\BibitemShut {NoStop}%
\bibitem [{\citenamefont {Yoshida}\ \emph {et~al.}(2024)\citenamefont {Yoshida}, \citenamefont {Zhang}, \citenamefont {Neupert},\ and\ \citenamefont {Kawakami}}]{prl2024norio076502}%
  \BibitemOpen
  \bibfield  {author} {\bibinfo {author} {\bibfnamefont {T.}~\bibnamefont {Yoshida}}, \bibinfo {author} {\bibfnamefont {S.-B.}\ \bibnamefont {Zhang}}, \bibinfo {author} {\bibfnamefont {T.}~\bibnamefont {Neupert}},\ and\ \bibinfo {author} {\bibfnamefont {N.}~\bibnamefont {Kawakami}},\ }\bibfield  {title} {\bibinfo {title} {Non-hermitian mott skin effect},\ }\href {https://doi.org/10.1103/PhysRevLett.133.076502} {\bibfield  {journal} {\bibinfo  {journal} {Phys. Rev. Lett.}\ }\textbf {\bibinfo {volume} {133}},\ \bibinfo {pages} {076502} (\bibinfo {year} {2024})}\BibitemShut {NoStop}%
\bibitem [{\citenamefont {Zhao}\ \emph {et~al.}(2025)\citenamefont {Zhao}, \citenamefont {Wang}, \citenamefont {He}, \citenamefont {Poon}, \citenamefont {Pak}, \citenamefont {Liu}, \citenamefont {Ren}, \citenamefont {Liu},\ and\ \citenamefont {Jo}}]{zhaoTwodimensionalNonHermitianSkin2025a}%
  \BibitemOpen
  \bibfield  {author} {\bibinfo {author} {\bibfnamefont {E.}~\bibnamefont {Zhao}}, \bibinfo {author} {\bibfnamefont {Z.}~\bibnamefont {Wang}}, \bibinfo {author} {\bibfnamefont {C.}~\bibnamefont {He}}, \bibinfo {author} {\bibfnamefont {T.~F.~J.}\ \bibnamefont {Poon}}, \bibinfo {author} {\bibfnamefont {K.~K.}\ \bibnamefont {Pak}}, \bibinfo {author} {\bibfnamefont {Y.-J.}\ \bibnamefont {Liu}}, \bibinfo {author} {\bibfnamefont {P.}~\bibnamefont {Ren}}, \bibinfo {author} {\bibfnamefont {X.-J.}\ \bibnamefont {Liu}},\ and\ \bibinfo {author} {\bibfnamefont {G.-B.}\ \bibnamefont {Jo}},\ }\bibfield  {title} {\bibinfo {title} {Two-dimensional non-{{Hermitian}} skin effect in an ultracold {{Fermi}} gas},\ }\href {https://doi.org/10.1038/s41586-024-08347-3} {\bibfield  {journal} {\bibinfo  {journal} {Nature}\ }\textbf {\bibinfo {volume} {637}},\ \bibinfo {pages} {565} (\bibinfo {year} {2025})}\BibitemShut {NoStop}%
\bibitem [{\citenamefont {Lee}\ \emph {et~al.}(2019)\citenamefont {Lee}, \citenamefont {Li},\ and\ \citenamefont {Gong}}]{lee_hybrid_2019}%
  \BibitemOpen
  \bibfield  {author} {\bibinfo {author} {\bibfnamefont {C.~H.}\ \bibnamefont {Lee}}, \bibinfo {author} {\bibfnamefont {L.}~\bibnamefont {Li}},\ and\ \bibinfo {author} {\bibfnamefont {J.}~\bibnamefont {Gong}},\ }\bibfield  {title} {\bibinfo {title} {Hybrid higher-order skin-topological modes in nonreciprocal systems},\ }\href {https://doi.org/10.1103/PhysRevLett.123.016805} {\bibfield  {journal} {\bibinfo  {journal} {Phys. Rev. Lett.}\ }\textbf {\bibinfo {volume} {123}},\ \bibinfo {pages} {016805} (\bibinfo {year} {2019})}\BibitemShut {NoStop}%
\bibitem [{\citenamefont {Li}\ \emph {et~al.}(2020{\natexlab{b}})\citenamefont {Li}, \citenamefont {Lee},\ and\ \citenamefont {Gong}}]{li_topological_2020}%
  \BibitemOpen
  \bibfield  {author} {\bibinfo {author} {\bibfnamefont {L.}~\bibnamefont {Li}}, \bibinfo {author} {\bibfnamefont {C.~H.}\ \bibnamefont {Lee}},\ and\ \bibinfo {author} {\bibfnamefont {J.}~\bibnamefont {Gong}},\ }\bibfield  {title} {\bibinfo {title} {Topological switch for non-{H}ermitian skin effect in cold-atom systems with loss},\ }\href {https://doi.org/10.1103/PhysRevLett.124.250402} {\bibfield  {journal} {\bibinfo  {journal} {Phys. Rev. Lett.}\ }\textbf {\bibinfo {volume} {124}},\ \bibinfo {pages} {250402} (\bibinfo {year} {2020}{\natexlab{b}})}\BibitemShut {NoStop}%
\bibitem [{\citenamefont {Kawabata}\ \emph {et~al.}(2020)\citenamefont {Kawabata}, \citenamefont {Sato},\ and\ \citenamefont {Shiozaki}}]{kawabata_higherorder_2020}%
  \BibitemOpen
  \bibfield  {author} {\bibinfo {author} {\bibfnamefont {K.}~\bibnamefont {Kawabata}}, \bibinfo {author} {\bibfnamefont {M.}~\bibnamefont {Sato}},\ and\ \bibinfo {author} {\bibfnamefont {K.}~\bibnamefont {Shiozaki}},\ }\bibfield  {title} {\bibinfo {title} {Higher-order non-{H}ermitian skin effect},\ }\href {https://doi.org/10.1103/PhysRevB.102.205118} {\bibfield  {journal} {\bibinfo  {journal} {Phys. Rev. B}\ }\textbf {\bibinfo {volume} {102}},\ \bibinfo {pages} {205118} (\bibinfo {year} {2020})}\BibitemShut {NoStop}%
\bibitem [{\citenamefont {Zou}\ \emph {et~al.}(2021)\citenamefont {Zou}, \citenamefont {Chen}, \citenamefont {He}, \citenamefont {Bao}, \citenamefont {Lee}, \citenamefont {Sun},\ and\ \citenamefont {Zhang}}]{zou_observation_2021}%
  \BibitemOpen
  \bibfield  {author} {\bibinfo {author} {\bibfnamefont {D.}~\bibnamefont {Zou}}, \bibinfo {author} {\bibfnamefont {T.}~\bibnamefont {Chen}}, \bibinfo {author} {\bibfnamefont {W.}~\bibnamefont {He}}, \bibinfo {author} {\bibfnamefont {J.}~\bibnamefont {Bao}}, \bibinfo {author} {\bibfnamefont {C.~H.}\ \bibnamefont {Lee}}, \bibinfo {author} {\bibfnamefont {H.}~\bibnamefont {Sun}},\ and\ \bibinfo {author} {\bibfnamefont {X.}~\bibnamefont {Zhang}},\ }\bibfield  {title} {\bibinfo {title} {Observation of hybrid higher-order skin-topological effect in non-{H}ermitian topolectrical circuits},\ }\href {https://doi.org/10.1038/s41467-021-26414-5} {\bibfield  {journal} {\bibinfo  {journal} {Nat. Commun.}\ }\textbf {\bibinfo {volume} {12}},\ \bibinfo {pages} {7201} (\bibinfo {year} {2021})}\BibitemShut {NoStop}%
\bibitem [{\citenamefont {Fu}\ \emph {et~al.}(2021)\citenamefont {Fu}, \citenamefont {Hu},\ and\ \citenamefont {Wan}}]{fu_nonhermitian_2021}%
  \BibitemOpen
  \bibfield  {author} {\bibinfo {author} {\bibfnamefont {Y.}~\bibnamefont {Fu}}, \bibinfo {author} {\bibfnamefont {J.}~\bibnamefont {Hu}},\ and\ \bibinfo {author} {\bibfnamefont {S.}~\bibnamefont {Wan}},\ }\bibfield  {title} {\bibinfo {title} {Non-{H}ermitian second-order skin and topological modes},\ }\href {https://doi.org/10.1103/PhysRevB.103.045420} {\bibfield  {journal} {\bibinfo  {journal} {Phys. Rev. B}\ }\textbf {\bibinfo {volume} {103}},\ \bibinfo {pages} {045420} (\bibinfo {year} {2021})}\BibitemShut {NoStop}%
\bibitem [{\citenamefont {Zhu}\ and\ \citenamefont {Gong}(2022)}]{zhu_hybrid_2022}%
  \BibitemOpen
  \bibfield  {author} {\bibinfo {author} {\bibfnamefont {W.}~\bibnamefont {Zhu}}\ and\ \bibinfo {author} {\bibfnamefont {J.}~\bibnamefont {Gong}},\ }\bibfield  {title} {\bibinfo {title} {Hybrid skin-topological modes without asymmetric couplings},\ }\href {https://doi.org/10.1103/PhysRevB.106.035425} {\bibfield  {journal} {\bibinfo  {journal} {Phys. Rev. B}\ }\textbf {\bibinfo {volume} {106}},\ \bibinfo {pages} {035425} (\bibinfo {year} {2022})}\BibitemShut {NoStop}%
\bibitem [{\citenamefont {Li}\ \emph {et~al.}(2022{\natexlab{b}})\citenamefont {Li}, \citenamefont {Liang}, \citenamefont {Wang}, \citenamefont {Lu},\ and\ \citenamefont {Liu}}]{li_gainlossinduced_2022}%
  \BibitemOpen
  \bibfield  {author} {\bibinfo {author} {\bibfnamefont {Y.}~\bibnamefont {Li}}, \bibinfo {author} {\bibfnamefont {C.}~\bibnamefont {Liang}}, \bibinfo {author} {\bibfnamefont {C.}~\bibnamefont {Wang}}, \bibinfo {author} {\bibfnamefont {C.}~\bibnamefont {Lu}},\ and\ \bibinfo {author} {\bibfnamefont {Y.-C.}\ \bibnamefont {Liu}},\ }\bibfield  {title} {\bibinfo {title} {Gain-loss-induced hybrid skin-topological effect},\ }\href {https://doi.org/10.1103/PhysRevLett.128.223903} {\bibfield  {journal} {\bibinfo  {journal} {Phys. Rev. Lett.}\ }\textbf {\bibinfo {volume} {128}},\ \bibinfo {pages} {223903} (\bibinfo {year} {2022}{\natexlab{b}})}\BibitemShut {NoStop}%
\bibitem [{\citenamefont {Sun}\ \emph {et~al.}(2023)\citenamefont {Sun}, \citenamefont {Li}, \citenamefont {Feng},\ and\ \citenamefont {Guo}}]{sun_hybrid_2023}%
  \BibitemOpen
  \bibfield  {author} {\bibinfo {author} {\bibfnamefont {J.}~\bibnamefont {Sun}}, \bibinfo {author} {\bibfnamefont {C.-A.}\ \bibnamefont {Li}}, \bibinfo {author} {\bibfnamefont {S.}~\bibnamefont {Feng}},\ and\ \bibinfo {author} {\bibfnamefont {H.}~\bibnamefont {Guo}},\ }\bibfield  {title} {\bibinfo {title} {Hybrid higher-order skin-topological effect in hyperbolic lattices},\ }\href {https://doi.org/10.1103/PhysRevB.108.075122} {\bibfield  {journal} {\bibinfo  {journal} {Phys. Rev. B}\ }\textbf {\bibinfo {volume} {108}},\ \bibinfo {pages} {075122} (\bibinfo {year} {2023})}\BibitemShut {NoStop}%
\bibitem [{\citenamefont {Sun}\ \emph {et~al.}(2024)\citenamefont {Sun}, \citenamefont {Hou}, \citenamefont {Wan}, \citenamefont {Wang}, \citenamefont {Zhu}, \citenamefont {Ruan},\ and\ \citenamefont {Yang}}]{sun_photonic_2024}%
  \BibitemOpen
  \bibfield  {author} {\bibinfo {author} {\bibfnamefont {Y.}~\bibnamefont {Sun}}, \bibinfo {author} {\bibfnamefont {X.}~\bibnamefont {Hou}}, \bibinfo {author} {\bibfnamefont {T.}~\bibnamefont {Wan}}, \bibinfo {author} {\bibfnamefont {F.}~\bibnamefont {Wang}}, \bibinfo {author} {\bibfnamefont {S.}~\bibnamefont {Zhu}}, \bibinfo {author} {\bibfnamefont {Z.}~\bibnamefont {Ruan}},\ and\ \bibinfo {author} {\bibfnamefont {Z.}~\bibnamefont {Yang}},\ }\bibfield  {title} {\bibinfo {title} {Photonic {F}loquet skin-topological effect},\ }\href {https://doi.org/10.1103/PhysRevLett.132.063804} {\bibfield  {journal} {\bibinfo  {journal} {Phys. Rev. Lett.}\ }\textbf {\bibinfo {volume} {132}},\ \bibinfo {pages} {063804} (\bibinfo {year} {2024})}\BibitemShut {NoStop}%
\bibitem [{\citenamefont {Ji}\ \emph {et~al.}(2024)\citenamefont {Ji}, \citenamefont {Ding}, \citenamefont {Chen},\ and\ \citenamefont {Yang}}]{ji_nonhermitian_2024}%
  \BibitemOpen
  \bibfield  {author} {\bibinfo {author} {\bibfnamefont {X.}~\bibnamefont {Ji}}, \bibinfo {author} {\bibfnamefont {W.}~\bibnamefont {Ding}}, \bibinfo {author} {\bibfnamefont {Y.}~\bibnamefont {Chen}},\ and\ \bibinfo {author} {\bibfnamefont {X.}~\bibnamefont {Yang}},\ }\bibfield  {title} {\bibinfo {title} {Non-{H}ermitian second-order topological superconductors},\ }\href {https://doi.org/10.1103/PhysRevB.109.125420} {\bibfield  {journal} {\bibinfo  {journal} {Phys. Rev. B}\ }\textbf {\bibinfo {volume} {109}},\ \bibinfo {pages} {125420} (\bibinfo {year} {2024})}\BibitemShut {NoStop}%
\bibitem [{\citenamefont {Zhu}\ and\ \citenamefont {Li}(2024)}]{JPCM2024_lhli}%
  \BibitemOpen
  \bibfield  {author} {\bibinfo {author} {\bibfnamefont {W.}~\bibnamefont {Zhu}}\ and\ \bibinfo {author} {\bibfnamefont {L.}~\bibnamefont {Li}},\ }\bibfield  {title} {\bibinfo {title} {A brief review of hybrid skin-topological effect},\ }\href {https://doi.org/10.1088/1361-648X/ad3593} {\bibfield  {journal} {\bibinfo  {journal} {J. Phys. Condens. Matter}\ }\textbf {\bibinfo {volume} {36}},\ \bibinfo {pages} {253003} (\bibinfo {year} {2024})}\BibitemShut {NoStop}%
\bibitem [{\citenamefont {Dash}\ \emph {et~al.}(2025)\citenamefont {Dash}, \citenamefont {Panda},\ and\ \citenamefont {Nandy}}]{prb2025dash}%
  \BibitemOpen
  \bibfield  {author} {\bibinfo {author} {\bibfnamefont {G.~K.}\ \bibnamefont {Dash}}, \bibinfo {author} {\bibfnamefont {S.}~\bibnamefont {Panda}},\ and\ \bibinfo {author} {\bibfnamefont {S.}~\bibnamefont {Nandy}},\ }\bibfield  {title} {\bibinfo {title} {Fingerprint of non-{{Hermiticity}} in a d -wave altermagnet},\ }\href {https://doi.org/10.1103/PhysRevB.111.155119} {\bibfield  {journal} {\bibinfo  {journal} {Phys. Rev. B}\ }\textbf {\bibinfo {volume} {111}},\ \bibinfo {pages} {155119} (\bibinfo {year} {2025})}\BibitemShut {NoStop}%
\bibitem [{\citenamefont {Reja}\ and\ \citenamefont {Narayan}(2024)}]{prb2024reja}%
  \BibitemOpen
  \bibfield  {author} {\bibinfo {author} {\bibfnamefont {M.~A.}\ \bibnamefont {Reja}}\ and\ \bibinfo {author} {\bibfnamefont {A.}~\bibnamefont {Narayan}},\ }\bibfield  {title} {\bibinfo {title} {Emergence of tunable exceptional points in altermagnet-ferromagnet junctions},\ }\href {https://doi.org/10.1103/PhysRevB.110.235401} {\bibfield  {journal} {\bibinfo  {journal} {Phys. Rev. B}\ }\textbf {\bibinfo {volume} {110}},\ \bibinfo {pages} {235401} (\bibinfo {year} {2024})}\BibitemShut {NoStop}%
\bibitem [{\citenamefont {Bernevig}\ \emph {et~al.}(2006)\citenamefont {Bernevig}, \citenamefont {Hughes},\ and\ \citenamefont {Zhang}}]{science2006sczhang}%
  \BibitemOpen
  \bibfield  {author} {\bibinfo {author} {\bibfnamefont {B.~A.}\ \bibnamefont {Bernevig}}, \bibinfo {author} {\bibfnamefont {T.~L.}\ \bibnamefont {Hughes}},\ and\ \bibinfo {author} {\bibfnamefont {S.-C.}\ \bibnamefont {Zhang}},\ }\bibfield  {title} {\bibinfo {title} {Quantum spin hall effect and topological phase transition in hgte quantum wells},\ }\href {https://doi.org/10.1126/science.1133734} {\bibfield  {journal} {\bibinfo  {journal} {Science}\ }\textbf {\bibinfo {volume} {314}},\ \bibinfo {pages} {1757} (\bibinfo {year} {2006})}\BibitemShut {NoStop}%
\bibitem [{\citenamefont {Yan}(2019)}]{yan_majorana_2019}%
  \BibitemOpen
  \bibfield  {author} {\bibinfo {author} {\bibfnamefont {Z.}~\bibnamefont {Yan}},\ }\bibfield  {title} {\bibinfo {title} {Majorana corner and hinge modes in second-order topological insulator/superconductor heterostructures},\ }\href {https://doi.org/10.1103/PhysRevB.100.205406} {\bibfield  {journal} {\bibinfo  {journal} {Phys. Rev. B}\ }\textbf {\bibinfo {volume} {100}},\ \bibinfo {pages} {205406} (\bibinfo {year} {2019})}\BibitemShut {NoStop}%
\bibitem [{\citenamefont {Qian}\ \emph {et~al.}(2014)\citenamefont {Qian}, \citenamefont {Liu}, \citenamefont {Fu},\ and\ \citenamefont {Li}}]{science2014jli}%
  \BibitemOpen
  \bibfield  {author} {\bibinfo {author} {\bibfnamefont {X.}~\bibnamefont {Qian}}, \bibinfo {author} {\bibfnamefont {J.}~\bibnamefont {Liu}}, \bibinfo {author} {\bibfnamefont {L.}~\bibnamefont {Fu}},\ and\ \bibinfo {author} {\bibfnamefont {J.}~\bibnamefont {Li}},\ }\bibfield  {title} {\bibinfo {title} {Quantum spin hall effect in two-dimensional transition metal dichalcogenides},\ }\href {https://doi.org/10.1126/science.1256815} {\bibfield  {journal} {\bibinfo  {journal} {Science}\ }\textbf {\bibinfo {volume} {346}},\ \bibinfo {pages} {1344} (\bibinfo {year} {2014})}\BibitemShut {NoStop}%
\bibitem [{\citenamefont {Ghorashi}\ \emph {et~al.}(2024{\natexlab{b}})\citenamefont {Ghorashi}, \citenamefont {Hughes},\ and\ \citenamefont {Cano}}]{ghorashi_altermagnetic_2024}%
  \BibitemOpen
  \bibfield  {author} {\bibinfo {author} {\bibfnamefont {S.~A.~A.}\ \bibnamefont {Ghorashi}}, \bibinfo {author} {\bibfnamefont {T.~L.}\ \bibnamefont {Hughes}},\ and\ \bibinfo {author} {\bibfnamefont {J.}~\bibnamefont {Cano}},\ }\bibfield  {title} {\bibinfo {title} {Altermagnetic routes to majorana modes in zero net magnetization},\ }\href {https://doi.org/10.1103/PhysRevLett.133.106601} {\bibfield  {journal} {\bibinfo  {journal} {Phys. Rev. Lett.}\ }\textbf {\bibinfo {volume} {133}},\ \bibinfo {pages} {106601} (\bibinfo {year} {2024}{\natexlab{b}})}\BibitemShut {NoStop}%
\bibitem [{\citenamefont {Liang}\ \emph {et~al.}(2022)\citenamefont {Liang}, \citenamefont {Xie}, \citenamefont {Dong}, \citenamefont {Li}, \citenamefont {Li}, \citenamefont {Gadway}, \citenamefont {Yi},\ and\ \citenamefont {Yan}}]{liang_dynamic_2022}%
  \BibitemOpen
  \bibfield  {author} {\bibinfo {author} {\bibfnamefont {Q.}~\bibnamefont {Liang}}, \bibinfo {author} {\bibfnamefont {D.}~\bibnamefont {Xie}}, \bibinfo {author} {\bibfnamefont {Z.}~\bibnamefont {Dong}}, \bibinfo {author} {\bibfnamefont {H.}~\bibnamefont {Li}}, \bibinfo {author} {\bibfnamefont {H.}~\bibnamefont {Li}}, \bibinfo {author} {\bibfnamefont {B.}~\bibnamefont {Gadway}}, \bibinfo {author} {\bibfnamefont {W.}~\bibnamefont {Yi}},\ and\ \bibinfo {author} {\bibfnamefont {B.}~\bibnamefont {Yan}},\ }\bibfield  {title} {\bibinfo {title} {Dynamic signatures of non-{H}ermitian skin effect and topology in ultracold atoms},\ }\href {https://doi.org/10.1103/PhysRevLett.129.070401} {\bibfield  {journal} {\bibinfo  {journal} {Phys. Rev. Lett.}\ }\textbf {\bibinfo {volume} {129}},\ \bibinfo {pages} {070401} (\bibinfo {year} {2022})}\BibitemShut {NoStop}%
\bibitem [{\citenamefont {Zhu}\ \emph {et~al.}(2018)\citenamefont {Zhu}, \citenamefont {Fang}, \citenamefont {Li}, \citenamefont {Sun}, \citenamefont {Li}, \citenamefont {Jing},\ and\ \citenamefont {Chen}}]{weiweiPhysRevLettacous}%
  \BibitemOpen
  \bibfield  {author} {\bibinfo {author} {\bibfnamefont {W.}~\bibnamefont {Zhu}}, \bibinfo {author} {\bibfnamefont {X.}~\bibnamefont {Fang}}, \bibinfo {author} {\bibfnamefont {D.}~\bibnamefont {Li}}, \bibinfo {author} {\bibfnamefont {Y.}~\bibnamefont {Sun}}, \bibinfo {author} {\bibfnamefont {Y.}~\bibnamefont {Li}}, \bibinfo {author} {\bibfnamefont {Y.}~\bibnamefont {Jing}},\ and\ \bibinfo {author} {\bibfnamefont {H.}~\bibnamefont {Chen}},\ }\bibfield  {title} {\bibinfo {title} {Simultaneous observation of a topological edge state and exceptional point in an open and non-hermitian acoustic system},\ }\href {https://doi.org/10.1103/PhysRevLett.121.124501} {\bibfield  {journal} {\bibinfo  {journal} {Phys. Rev. Lett.}\ }\textbf {\bibinfo {volume} {121}},\ \bibinfo {pages} {124501} (\bibinfo {year} {2018})}\BibitemShut {NoStop}%
\bibitem [{\citenamefont {Helbig}\ \emph {et~al.}(2020)\citenamefont {Helbig}, \citenamefont {Hofmann}, \citenamefont {Imhof}, \citenamefont {Abdelghany}, \citenamefont {Kiessling}, \citenamefont {Molenkamp}, \citenamefont {Lee}, \citenamefont {Szameit}, \citenamefont {Greiter},\ and\ \citenamefont {Thomale}}]{helbig_generalized_2020}%
  \BibitemOpen
  \bibfield  {author} {\bibinfo {author} {\bibfnamefont {T.}~\bibnamefont {Helbig}}, \bibinfo {author} {\bibfnamefont {T.}~\bibnamefont {Hofmann}}, \bibinfo {author} {\bibfnamefont {S.}~\bibnamefont {Imhof}}, \bibinfo {author} {\bibfnamefont {M.}~\bibnamefont {Abdelghany}}, \bibinfo {author} {\bibfnamefont {T.}~\bibnamefont {Kiessling}}, \bibinfo {author} {\bibfnamefont {{\relax LW}.}~\bibnamefont {Molenkamp}}, \bibinfo {author} {\bibfnamefont {{\relax CH}.}~\bibnamefont {Lee}}, \bibinfo {author} {\bibfnamefont {A.}~\bibnamefont {Szameit}}, \bibinfo {author} {\bibfnamefont {M.}~\bibnamefont {Greiter}},\ and\ \bibinfo {author} {\bibfnamefont {R.}~\bibnamefont {Thomale}},\ }\bibfield  {title} {\bibinfo {title} {Generalized bulk--boundary correspondence in non-{H}ermitian topolectrical circuits},\ }\href {https://doi.org/10.1038/s41567-020-0922-9} {\bibfield  {journal} {\bibinfo  {journal} {Nat. Phys.}\ }\textbf {\bibinfo {volume} {16}},\ \bibinfo {pages} {747} (\bibinfo {year} {2020})}\BibitemShut {NoStop}%
\end{thebibliography}%
\end{document}